\documentclass[submission,copyright,creativecommons,noderivs,noncommercial]{eptcs}
\providecommand{\event}{Seventh International Workshop on Formal\\ Methods for Autonomous Systems}
\usepackage{iftex}

\usepackage[utf8]{inputenc}
\usepackage[english]{babel}
\usepackage{amssymb}
\usepackage{amsmath}
\usepackage{amsthm}
\newtheorem{definition}{Definition}
\theoremstyle{definition}
\newtheorem{characterization}{Characterisation}
\newtheorem{example}{Example}

\usepackage{todonotes}
\usepackage{csquotes}
\usepackage[per-mode=symbol, binary-units=true]{siunitx}
\usepackage{caption}
\usepackage{graphicx}
\usepackage{subcaption}
\usepackage{caption}
\usepackage{wrapfig}
\usepackage{csquotes}
\usepackage{makecell}
\usepackage{fdsymbol}
\usepackage{float}
\usepackage{lipsum}
\usepackage{booktabs}
\usepackage{xspace}
\usepackage{xcolor}
\edef\restoreparindent{\parindent=\the\parindent\relax}
\newcommand{\doi}[1]{\textsc{doi}: \href{http://dx.doi.org/#1}{\nolinkurl{#1}}}
\restoreparindent
\usepackage{tsc}
\usetikzlibrary{arrows}
\usetikzlibrary{trees,shapes,decorations,fit}
\usetikzlibrary{shapes.multipart,arrows.meta}
\usepackage{lipsum}  
\tikzset{%
	>={Latex[width=2mm,length=2mm]},
	base/.style = {rectangle, rounded corners, draw=black,
		minimum width=4cm, minimum height=1.5cm,
		text centered, font=\footnotesize},
	activityStarts/.style = {base, fill=blue!30},
	inner/.style = {base, dashed, line width=1.5pt},
	activityRuns/.style = {base, draw=green!70!black, line width=1.5pt},
	process/.style = {base, minimum width=2.5cm, line width=1.5pt},
}

\DeclareMathAlphabet{\mathpzc}{OT1}{pzc}{m}{it}

\addto\extrasenglish{%
	\renewcommand{\sectionautorefname}{Section}%
}



\newcommand{\Autoref}[1]{%
	\begingroup%
	\renewcommand\equationautorefname{Equation}%
	\renewcommand\figureautorefname{Figure}%
	\renewcommand\subsectionautorefname{Section}%
	\renewcommand\subsubsectionautorefname{Section}%
	\renewcommand\sectionautorefname{Section}%
	\renewcommand\tableautorefname{Table}%
	\autoref{#1}%
	\endgroup%
}

\def\subsubsectionautorefname{Sec.}
\def\subsectionautorefname{Sec.}
\def\sectionautorefname{Sec.}
\def\tableautorefname{Tab.}
\def\figureautorefname{Fig.}
\def\equationautorefname{Eq.}

\graphicspath{{figures_/}}

\usepackage{todonotes}

\usepackage{adjustbox}

\newcommand{\AVs}{\textit{AVs}\xspace}
\newcommand{\AV}{\textit{AV}\xspace}

\newcommand{\FE}{\mathcal{F}_E}
\newcommand{\FV}{\mathcal{F}_V}

\newcommand{\Ss}{\mathbf{S}}
\usepackage{array}
\newcolumntype{P}[1]{>{\centering\arraybackslash}p{#1}}
\newcolumntype{M}[1]{>{\centering\arraybackslash}m{#1}}

\usepackage[usestackEOL]{stackengine}
\usepackage[inline]{enumitem}

\title{Context-aware, Ante-hoc Explanations of Driving Behaviour}
\author{Dominik Grundt
\and
\hspace*{-1em}Ishan Saxena
\and
\hspace*{-1em}Malte Petersen
\and
\hspace*{-1em}Bernd Westphal
\email{\hspace*{-18em}<firstname.lastname>@dlr.de}
\and
\hspace*{-1em}Eike M\"ohlmann
}

\def\titlerunning{Context-aware, Ante-hoc Explanations of Driving Behaviour}
\def\authorrunning{D. Grundt, I. Saxena, M. Petersen, B. Westphal, E. M\"ohlmann}
\def\copyrightholders{Grundt, Saxena, Petersen, Westphal \& M\"ohlmann}
\begin{document}
\maketitle
\begin{center}
	\vspace*{-1em}
	{\small German Aerospace Center \\ 
	Institute of Systems Engineering for Future Mobility \\
	Oldenburg, Germany}
\end{center}

\vspace*{.3cm}
\begin{abstract}
Autonomous vehicles (\AVs) must be both safe and trustworthy to gain social acceptance 
and become a viable option for everyday public transportation. 
Explanations about the system behaviour can increase safety and trust in \AVs.  
Unfortunately, explaining the system behaviour of AI-based driving functions is particularly challenging, as decision-making processes are often opaque. The field of \textit{Explainability Engineering} tackles this challenge by developing \textit{explanation models} at design time. These models are designed from system design artefacts and stakeholder needs to develop \textit{correct} and \textit{good} explanations.
To support this field, we propose an approach that enables context-aware, ante-hoc explanations of (un)expectable driving manoeuvres at runtime. 
The visual yet formal language Traffic Sequence Charts is used to formalise \textit{explanation contexts}, as well as corresponding (un)expectable driving manoeuvres. A dedicated runtime monitoring enables context-recognition and ante-hoc presentation of explanations at runtime. In combination, we aim to support the bridging of \textit{correct} and \textit{good} explanations. Our method is demonstrated in a simulated overtaking.
\end{abstract}
\thanks{\footnotesize{\noindent%
The research leading to these results is funded by the German Federal Ministry of Education and Research under grant agreement No~16MEE044 (EdgeAI-Trust) and
by the Chips Joint Undertaking under grant agreement No~101139892 (EdgeAI-Trust).}}
\section{Introduction}
\label{sec:intro}


To deploy autonomous vehicles (\AVs) in public road traffic and to gain societal acceptance, they must be safe and trustworthy. Achieving these properties in road traffic is difficult for two main reasons. First, the inherently dynamic and open-world nature of road traffic introduces uncertainty, as \AVs must operate under constantly changing and only partially predictable conditions. Second, the black-box character of many AI-based solutions limits transparency and interpretability, making it difficult to assess their correctness, especially crucial in safety-critical situations.  
In combination, these challenges make the specification of a complete formal behaviour model of \AVs infeasible.

On the regulatory level, the upcoming European AI Act~\cite{euaiact} mandates transparency, human oversight, and traceability for high-risk AI systems, including \AVs. Moreover, studies such as \cite{Othman2021} and \cite{Alqahtani2025} reveal that social acceptance of \AVs remains limited, mainly due to a lack of trust. Consequently, there is an urgent need for reliable explanation methods in autonomous driving, even in the absence of a complete formal behavioural model.

The field of Explainability Engineering addresses this challenge by developing methods and tools to systematically design explanation models from system descriptions, stakeholder needs, and requirements at design time. The goal is to build a foundation for developing explanations that are both \textit{correct} and \textit{good}~\cite{Schwammberger2025}. Research in this field enhances trust, facilitates safety validation, and fulfils regulatory requirements by making AI decisions interpretable, traceable and understandable for different stakeholders.
Empirical research provides guidance on which explanations are effective in increasing trust.
Meta-studies indicate that while system performance is important, trust in autonomous systems is
significantly enhanced by behavioural explanations~\cite{Atf2025}.
Explaining \textit{expectable} manoeuvres helps stakeholders build confidence in an \AV’s reliable and plausible operation~\cite{Atf2025}.
Conversely, explaining \textit{unexpectable} manoeuvres is crucial for maintaining trust in safety-critical situations~\cite{Papagni2022}.
For this work, we refer to \textit{(un)expectable} driving manoeuvres as behaviour that is \textit{(not-)permitted} according to an \AV’s explanation model.
Rather than \textit{(un)expected}, the term \textit{(un)expectable} emphasises model-based admissibility instead of compliance with predefined requirements, while also capturing runtime uncertainties of the domain and AI-based driving functions.

Furthermore, explanations are most effective when provided \textit{before} an \AV executes a manoeuvre~\cite{Du2019}.
We refer to such explanations as \textit{ante-hoc}, meaning that information about (un)expectable behaviour is available prior to execution.
Prior work also shows that, to be effective, trustworthy, and meaningful, explanations of \AV behaviour must consider the surrounding traffic and environmental conditions~\cite{bohlender2019characterization,Kaufman2025,Ma2024}.
We refer to explanations that include such factors as \textit{context-aware}.

While frameworks such as MAB-EX~\cite{Blumreiter2019} describe how explainability can be achieved at runtime, and literature in Explainability Engineering defines properties and requirements for \textit{good} and \textit{correct} explanations~\cite{Schwammberger2025}, concrete methods to operationalise these concepts for \AVs are still lacking. In this work, we present a method to specify \textit{context-aware} explanations of (un)expectable driving manoeuvres and to provide such explanations \textit{ante-hoc} at runtime.

We use Traffic Sequence Charts (TSC)~\cite{atr117} to formally specify the \emph{explanation context} (or short \emph{context}), i.e., the traffic situation in which an explanation should be provided.
TSC allow the specification of spatio‑temporal properties, particularly of traffic scenarios, in both visual and formal form. Using TSC for context specification is beneficial for stakeholder needs, such as comprehensibility and goodness. For example, psychologists can assess both properties by investigation of the visual part of TSC without needing to understand the formal semantics.
Further, we use TSC runtime monitoring~\cite{Stemmer2025,Grundt2022} to recognise the specified explanation context at runtime and trigger the ante-hoc presentation of an explanation.
Finally, we show how (un)expectable driving manoeuvres can be specified with TSC. 

Therefore, we address the challenges of (i) formally specifying explanation contexts for \AVs, (ii) enabling ante-hoc explanations at runtime, and (iii) enabling different forms of an explanation about (un)expectable driving manoeuvres.

\textit{Outline.} The paper is organised as follows: \Autoref{sec:preliminaries} provides preliminaries for our work. \Autoref{sec:rw} discusses related work. \Autoref{sec:tsc-explanations} presents the formalisation of an \textit{explanation context} using TSC. \Autoref{sec:monitoring} presents TSC runtime monitoring for context-recognition and ante-hoc explanations at runtime. \Autoref{sec:maneuver-spec} presents the visual yet formal specification of (un)expectable driving manoeuvres in a context-aware manner using TSC. \Autoref{sec:experiment} demonstrates our approach within a simulated overtaking. \Autoref{sec:conclusion} contains the conclusion and discusses future work.
\vspace*{-.25cm}
\section{Preliminaries}
\label{sec:preliminaries}

In this section, we give a brief introduction to the field of Explainability Engineering, including a characterisation of \textit{explanations} and describe how our approach aligns with the MAB-EX Framework~\cite{Blumreiter2019}.
%
%
%
\vspace*{-.9cm}
\paragraph{Explainability Engineering}\label{subsec:eeintro}
In \cite{Kohl2019}, explainability was characterised as a non-functional requirement. With its recent classification as a requirement also at the regulatory level, explainability has become
an engineering task for AI-based systems such as \AVs. Presenting pure information without context does not provide the necessary explainability that creates traceability, transparency and comprehensibility of system behaviour~\cite{Ma2024}. As shown in~\cite{Schwammberger2022EM}, stakeholder-specific explanation models for such complex systems need to be created systematically at design time. Explainability Engineering (EE) directly addresses this need for transparency and traceability at both regulatory~\cite{euaiact} and societal levels~\cite{Othman2021,Alqahtani2025} for high-risk AI systems.

While methods in eXplainable AI (XAI) primarily aim at generating explanations for trained, complex AI models, these explanations are typically tied to a specific model and often remain interpretable only to experts~\cite{Schwammberger2024xAI}. In contrast, EE pursues the broader goal of creating explanations for the behaviour of entire AI-based systems or systems of systems, such as \AVs.

An explanation model $EM$ may be defined as a behavioural model of the system that captures causal relationships between events and system reactions. This enables the identification of possible causes for behaviours that require explanation~\cite{Blumreiter2019}. Such models provide a conceptual foundation for delivering explanations to diverse stakeholders, from engineers needing technical traceability to passengers seeking intuitive justifications and regulators requiring formal evidence of compliance. In \cite{Schwammberger2022EM}, Schwammberger and Klös showed how to derive explanation models in the context of autonomous agents. In this paper, we assume the existence of an $EM$ that contains relations between traffic situations and corresponding (un)expectable driving manoeuvres.
\vspace*{-.35cm}
\paragraph{Explanations in Explainability Engineering}\label{subsec:explanations}
In our work, we use the following characterisation of explanations given by Schwammberger et al. as an extension of Bohlender and Köhl (see \cite{bohlender2019characterization}):
\begin{characterization}[Explanation~\cite{Schwammberger2024xAI}]\label{char:expl}
	\textit{An explanation $E$ is characterised by (i) explananda X, or “phenomena”, of the system of interest,
		(ii) a context $C$,
		(iii) a stakeholder group $G$ for $E$,
		(iv) the goal $\theta$ of $E$ for a stakeholder group $G$, and 
		(v) the means $M$ for producing $E$. }
\end{characterization}

In our approach, the explanandum $X$ of an explanation $E$ is which driving manoeuvres of an \AV are (un)expectable in the current context $C$, where the context represents the \AV’s current traffic situation.
Using TSC with dedicated runtime monitoring, both the context $C$ and the context-aware, ante-hoc presentation of explanations $E$ can be formalised in the same specification language.
Thus, for the explanation means $M$, we leverage the visual yet formal form of TSC to specify driving manoeuvres. 

This setup is deliberately modular. The TSC-based specification of context $C$ and its runtime monitoring can be replaced by any other context-recognition method. The TSC-specified driving manoeuvres continue to serve as the explanation means $M$. Conversely, the TSC-based context and monitoring can trigger context-aware, ante-hoc presentation of any other explanations.
Regarding our method, the goal $\theta$ is to increase trust in \AVs across different stakeholder groups $G$.
However, explanation \emph{goodness}~\cite{Schwammberger2025} needs to be investigated by future empirical studies.

\begin{wrapfigure}[11]{r}{0.3\textwidth}
	\centering
	\vspace*{-.3cm}
	\includegraphics[scale=0.3]{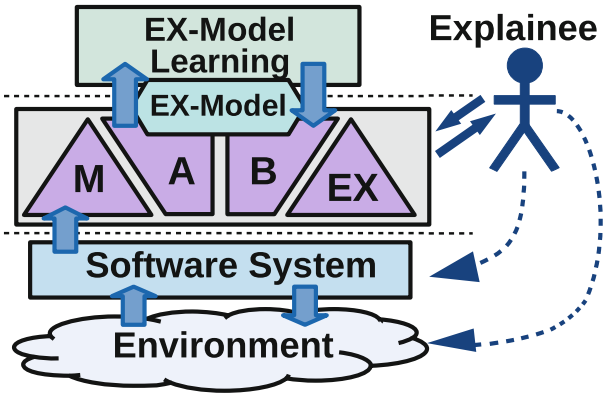} 
	\caption{MAB-EX Framework \cite{Blumreiter2019} (concretised for software systems in \cite{Schwammberger2024xAI}).}
	\label{fig:mabex}
\end{wrapfigure}

\vspace*{-.35cm}
\paragraph{MAB-EX Framework}\label{subsec:Mabex}
In EE, the \textit{MAB-EX} framework~\cite{Blumreiter2019} (depicted in \autoref{fig:mabex}) provides a theoretical
foundation for creating self-explanatory capabilities of systems at runtime. Our approach can be aligned with the four phases \textit{\textbf{M}onitoring}, \textit{\textbf{A}nalyse}, \textit{\textbf{B}uild}, and \textit{\textbf{EX}plain}. 

The system considered in this work is an $AV$ interacting with its environment. The explanation model captures the causal factors underlying the $AV$’s behaviour. Within MAB-EX, a learning mechanism for the explanation model is foreseen, which can iteratively improve the $AV$’s explanation model for different stakeholders using operational data from the dynamic and uncertain automotive domain. Both the design of the explanation model and any model learning remain out of scope. In this work, we primarily operationalise the four runtime phases for context-aware, ante-hoc explanations of (un)expectable driving manoeuvres.
The TSC runtime monitoring corresponds to the \textit{\textbf{M}onitoring} phase, where surrounding traffic and the environment are continuously observed. The subsequent context recognition, based on the TSC specification, maps to the \textit{\textbf{A}nalyse} phase, where the observed data are checked for the phenomenon to be explained (the explanandum $X$).
The TSC specification of traffic situations and corresponding (un)expectable manoeuvres, along with their runtime selection, belongs to the \textit{\textbf{B}uild} phase, where explanations are developed from an explanation model. Finally, the visual or formal presentation of (un)expectable manoeuvres corresponds to the \textit{\textbf{EX}plain} phase, communicating system behaviour in an interpretable and comprehensible way.

\vspace*{-.3cm}
\section{Related Work}\label{sec:rw}
Formal specification of spatio-temporal properties has already been used in the automotive domain to specify, e.g., system requirements based on languages such as LTL~\cite{Rizaldi2017,Esterle2020} and STL~\cite{Tuncali2018,Arechiga2019}, as well as for complex system requirements with TSC~\cite{Stemmer2025,Grundt2022}. Since no language has emerged as standard, we employ TSC for context and manoeuvrer specification, whose visual and formal syntax may be able to address multiple stakeholder groups. Visual or formal {\small XAI} presentations of explanations, such as saliency maps, heatmaps, or decision‑tree paths~\cite{Alicioglu2022}, target domain experts and require further interpretation. In contrast, TSC‑based specifications allow for deciding in advance what to communicate, how, and to whom, tailoring the explanation to the need, context and goals. This aligns with Explainability Engineering principles of designing \textit{correct} and \textit{good} explanations~\cite{Schwammberger2025}.

Runtime Monitoring (RM) is a technique used to check whether a system's runtime behaviour complies with formally specified properties~\cite{Bartocci2018IntroRV}. Particularly in safety-critical domains, RM ensures the correctness and safety of complex systems such as \AVs. RM continuously observes system behaviour to detect deviations from specifications~\cite{Bauer2011}, complementing offline verification and allowing timely identification of unsafe states~\cite{Havelund2002}. For this work, we use and adapt an already developed TSC runtime monitoring~\cite{Grundt2022,Stemmer2025} for recognising an explanation context specified using TSC. 

Schwammberger et al.~\cite{Schwammberger2022EM} demonstrate a possible derivation method for explanation models that considers different stakeholder groups and present a case study where a timed automaton models a crossing protocol for urban intersection turn manoeuvres. Their case study discusses the MAB-EX phases and explanation forms abstractly for multiple stakeholder groups. 
In a recent extension of the MAB-EX framework~\cite{Schwammberger2024xAI}, the authors introduce the concept of an \textit{explanation history} to support the recognition of recurring contexts and dynamically adapt explanations accordingly. This approach enables the generation of \textit{global} explanations that generalise across a variety of similar contexts.

This work can be seen as the first concrete operationalisation of MAB-EX runtime phases to provide context-aware, ante-hoc explanations of (un)expectable \AV manoeuvres at runtime. Complementing prior work on deriving explanation models for different stakeholder groups~\cite{Schwammberger2022EM} and on extending the MAB-EX framework with an explanation history~\cite{Schwammberger2024xAI}, our approach focuses on implementing the runtime phases in a context-aware manner. Together, these perspectives form a coherent research trajectory: stakeholder-specific explanation models provide the conceptual foundation, our operationalisation demonstrates runtime applicability, and extensions like explanation history enable adaptive refinement of explanations.

\section{Traffic Sequence Charts for Context-Aware Explanations}
\label{sec:tsc-explanations}
In the following, we show how TSC can be used to formalise an explanation context to enable context-recognition at runtime and specify context-aware explanations.

Recall that we assume the existence of an explanation model $EM$ that contains relations between traffic situations and corresponding (un-)expectable driving manoeuvres, created by system and domain experts at design time. 
Considering the characterisation of an explanation (see \autoref{char:expl}), we define the context $C$ of an explanation $E$ about the behaviour of an $\AV$ as the specification of a traffic situation.
A traffic situation of an $AV$ consists of the surrounding environment (e.g., lanes, markings, road signs) and traffic participants which interact with the environment. Traffic participants are constrained by behavioural and physical rules in the environment, and have attributes such as speed and position. Functional descriptions of traffic situations or a context $C$ for our explanation $E$ can be as follows:

\begin{example}[Context]\label{ex:context}
\textit{ $AV$ is driving towards a slow driving vehicle (slower than the speed limit) in its lane on a two-lane carriageway. There is no other vehicle on the adjacent left lane of $AV$. The distance between  $AV$ and the other vehicle is less than 25 metres.}
\end{example}
We refer to a concrete instance of context $C$, in which an explanation $E$ of $AV$ is to be presented, as a \textit{concrete traffic situation}. A concrete traffic situation captures the spatial relations between traffic participants and the environment at a specific point in time, including the fixed state of object attributes.

To formalise the concrete traffic situation as context $C$ of an explanation $E$, we model the traffic environment of $AV$, its traffic participants and their respective attributes in an object model $OM = (\mathcal{T},\mathcal{C},F,Pred)$, where $\mathcal{T}$ is a set of basic types, $\mathcal{C}$ is a set of object types (or classes), $F$ is a set of typed function symbols, and $Pred$ is a set of typed predicate symbols. With $attr(\mathbf{C})$, we describe the finite set of typed attributes for each object class $\mathbf{C} \in \mathcal{C}$.
Now, given an object model $OM$, we define a concrete traffic situation as a function 
$\sigma : ID \rightarrow attr(\mathbf{C}) \rightarrow \mathcal{D}$,
where $\mathbf{ID}$ denotes the finite set of object identities (i.e., the traffic objects involved in the situation), $attr(\mathbf{C})$ the attributes of the context, and $\mathcal{D}$ the domain of valid, type-consistent values.

In the following, we briefly introduce TSC and show how they can be used to formalise an explanation context $C$ as traffic situation specification. Such a specification characterises not a single, but a set of concrete traffic situations that an \AV might encounter, as in our example context.

\subsection{Traffic Sequence Charts}\label{subsec:tsc-intro}
TSC is a visual and formal language for specifying spatio-temporal properties in traffic environments. The simplest chart, a \textit{Basic Chart}, contains a single \textit{invariant node}. Each node encapsulates a \textit{Spatial View} (SV), the core visual formalism, expressing spatial relations (e.g., distances) and object-attribute predicates (e.g., velocity, acceleration), combined with Boolean operators.

Given the underlying time model $\mathbb{T}$ of TSC with continuous time semantics, any constraint holds for a non-empty time interval.
Therefore, a Basic Chart, i.e. an invariant node containing an SV, specifies propositional constraints that hold invariantly over an interval $[b,e] \subset \mathbb{T}$ with $e>b$.

TSC can be used to formalise the context $C$ as a traffic situation specification, i.e., a set of well-typed predicate logic formulae defined over an $OM$. Together with the aforementioned time model $\mathbb{T}$, TSC allow the specification of so-called abstract traffic situations $S$.

A so-called \textit{symbol dictionary} is essential for the visual syntax of SV. Object symbols from the symbol dictionary can be placed on a rectangular canvas, which exists within the corresponding invariant node. Each object symbol is assigned to exactly one object type $\mathbf{C}$ from $OM$ and may have unique logical object variables, often associated with an object's $id \in ID$. Anchors are always required and necessary for expressing spatial properties. Each object symbol has at least one anchor, e.g. in the centre of the symbol. Each anchor is linked to a position attribute $pos \in \mathbb{R}^2$ of an object type instance, which is always required. Exceptions are the so-called line anchors, which are fixed in one dimension on the 2-dimensional canvas. These can be used, for example, to logically specify the boundaries of a carriageway. Thus, by placing the symbols on the canvas, spatial relations of objects can be expressed via predicate logic formulae. The anchors form the basis for further elements such as distance lines (constraining the distance between two objects), nowhere-boxes (excluding object types in a specific area), or somewhere-boxes (defining a specific area in which an object can be). 
Predicate logic formulas for TSC specifications can be derived by the algorithms presented in \cite{atr117}. 
In the following, we provide a detailed description of both the visual and formal specification of the functional described traffic situation, i.e., our exemplary context (see \autoref{ex:context}).
\subsection{Specifying Traffic Situation $S_i \in \Ss$}\label{subsec:sit-spec}
\begin{figure}[h!]
	\centering
	\begin{subfigure}[b]{0.27\textwidth}
		\centering
		\includegraphics[width=\textwidth]{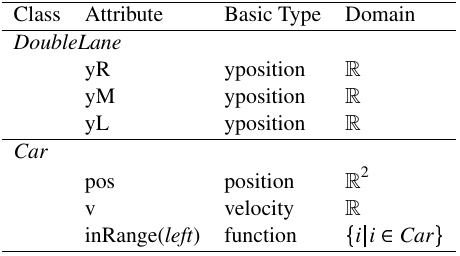}
		\caption{Object model $OM$ defining two classes, their attributes according to our example.}
		\label{fig:exOM}
	\end{subfigure}
	\hfill
	\begin{subfigure}[b]{0.28\textwidth}
		\centering
		\includegraphics[width=\textwidth]{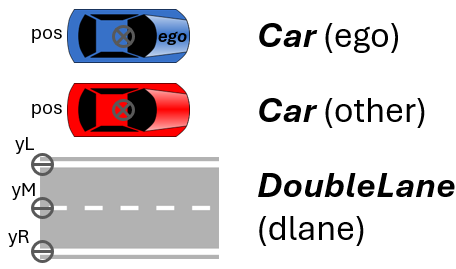}
		\caption{Symbol Dictionary defining object symbols and anchors according to our example.}
		\label{fig:exSD}
	\end{subfigure}
	\hfill
	\begin{subfigure}[b]{0.41\textwidth}
		\centering
		\includegraphics[width=0.8\textwidth]{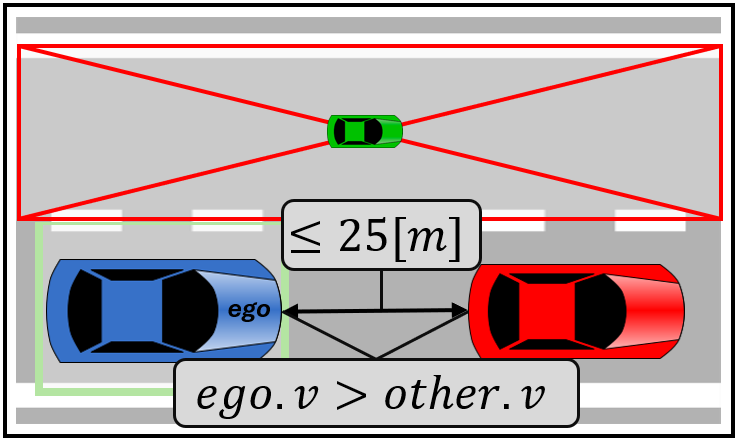}
		\caption{Basic Chart describing the spatial relations and constraints on object attributes according to our example.}
		\label{fig:exSV}
	\end{subfigure}
	\caption{TSC Basic Chart of the example with an object model $OM$ (a), a corresponding symbol dictionary (b), and an invariant node encapsulating one spatial view (c).}
	\label{fig:ex-tsc}
\end{figure}

The specification of our exemplary traffic situation as an abstract traffic situation using TSC is shown in \autoref{fig:ex-tsc}.
It defines the described traffic situation with a \textit{Basic Chart} (\autoref{fig:exSV}) over an $OM$ (\autoref{fig:exOM}) and a symbol dictionary (\autoref{fig:exSD}). The \AV (blue car symbol, named \textit{ego}) is placed left of the slower driving (constraint on velocity between both symbols) car (red car symbol, named \textit{other}) on the 2D canvas, expressing that \textit{ego} is driving towards \textit{other}. The described distance between \textit{ego} and \textit{other} is expressed by a \textit{distance line} connecting symbol anchors and annotated with the distance constraint.
Both car symbols have one anchor, while the two-lane road has three anchors for the three lane boundaries. This enables the specification of all described spatial relations.
A so-called somewhere-box around \textit{ego} and \textit{other} defines that any lateral position of the cars in their lane (between \textit{DoubleLane}'s anchors $yR$ and $yM$) is valid. A nowhere-box with a car symbol on the left lane (between \textit{DoubleLane}'s anchors $yM$ and $yL$) defines the described situation, that there is no other car on the adjacent left lane.

In addition, the $OM$ has a function called \textit{inRange()}. The nowhere-box can be mapped to this function so that \textit{ego} can determine how many other objects of the class \textit{Car} exist in the range (e.g., adjacent lane sensor range of a real car's lane change assistant (LCA)) of \textit{ego} in the left lane.

\begin{wrapfigure}[7]{r}{0.57\textwidth}
	\centering
	\vspace{-2.2em}
	\begin{subequations}\label{eq:exsv}
		\begin{align}
			B1 &= \, ego.pos.y > lane.yR \land ego.pos.y < lane.yM \label{eq:exsv_a}\\
			\land &\, other.pos.y > lane.yR \land other.pos.y < lane.yM \label{eq:exsv_b}\\
			\land &\, ego.pos.x + 25 \leq other.pos.x \label{eq:exsv_c}\\
			\land &\, ego.v > other.v \label{eq:exsv_d}\\
			\land &\, ego.inRange(left) = \emptyset \label{eq:exsv_e}
		\end{align}
	\end{subequations}
\end{wrapfigure}

The derived predicate logical formula (supported by the algorithms presented in \cite{atr117}) 
for the specification in \autoref{fig:ex-tsc} is shown in \autoref{eq:exsv}. In \autoref{eq:exsv_a}, the spatial relation between \textit{ego} and the \textit{DoubleLane} 
is defined over $OM$. 
In \autoref{eq:exsv_b} the spatial relation between the \textit{other} car and the \textit{DoubleLane} is defined. \autoref{eq:exsv_c} expresses the distance constraint between \textit{ego} and \textit{other}, and \autoref{eq:exsv_d} the difference in velocity.
\autoref{eq:exsv_e} expresses the constraint for the (car) empty adjacent left lane of \textit{ego}.

By specifying the exemplary traffic situation as an abstract traffic situation $S$ using TSC, a whole set of concrete situations is covered. Thus, in any concrete situation satisfying the abstract traffic situation, the same explanation will be applied. This allows for capturing infinitely many traffic situations with a finite number of explanations.

The formalisation of a traffic situation as TSC formalises the context $C$, which \textcolor{teal}{is} important for the interpretation, traceability, and comprehensibility of an explanation $E$. Such traffic situation specification $S$ also serves as the formal specification for the construction of a TSC runtime monitor. In \Autoref{sec:monitoring} we describe our use of TSC runtime monitoring for recognising the entering or being of $AV$ in a concrete traffic situation $\sigma$, which corresponds to traffic situation specification $S$, i.e., $\sigma \models S$. Hence, $S$ is a formal foundation for context-aware and ante-hoc presentation of any explanations for context $C$. 
\section{Context-Recognition and Ante-hoc Explanation at Runtime}
\label{sec:monitoring}
To enable context-aware, ante-hoc explanations of the behaviour of an $AV$, we need to recognise which concrete traffic situation $\sigma$ the $AV$ is currently in. The goal is to trigger the presentation of a context-aware explanation $E$, corresponding to the traffic situation $\sigma$. 
We use the term \textit{ante-hoc explanation} (of (un)expetable driving manoeuvres) to denote explanations that are presented before an $AV$ behaves as explained.
The term ante-hoc explanations is often used in the context of white-box AI models. However, here we use the term for explanations stemming from explanation models, that are designed using system design artefacts and stakeholder needs.
Providing such explanations requires rapid recognition of the \AV entering or being in a context $C$, as well as timely presentation. With this work, we contribute the technical capability for context-aware, ante-hoc explanations. Aspects such as optimal presentation duration or alignment with passengers’ cognitive load~\cite{Bairy2023} are beyond this work’s scope. In \Autoref{sec:experiment}, we demonstrate the feasibility of providing context-aware, ante-hoc explanations of (un)expectable manoeuvres at runtime.

In \Autoref{sec:tsc-explanations}, we showed how a described context $C$ of an explanation $E$ can be formalised as abstract traffic situations $S$ using TSC. Such a specification defines spatial relations and constraints for a set of object types, attributes, and predicates over a shared $OM$.

To determine whether an \AV is currently in a concrete traffic situation $\sigma$ corresponding to a TSC context specification $S$, we use runtime monitoring. A dedicated TSC runtime monitoring for scenario-based testing has been developed previously~\cite{Stemmer2025, Grundt2022}. Here, we adapt it to enable context-recognition and trigger ante-hoc explanations. Concretely, we check at runtime whether $\sigma \models S$ and, if so, present the corresponding explanation $E$.

\subsection{Matching Concrete Situations with Context Specifications}

Let $\sigma$ denote a concrete traffic situation, provided from sensor data at runtime. We assume the existence of a valuation function $\rho$ that maps each traffic object identity $id \in ID$ to its observed attribute values $\rho(id)(a)$, where $a \in attr(C)$ and $C \in OM$.
Given the specification of $S$ as an abstract traffic situation using TSC, such a specification can be interpreted as a set of well-typed logical predicates $P(S)$ over $OM$ (see \autoref{eq:exsv}). A concrete traffic situation $\sigma$ corresponds to a specification $S$ if
$
\sigma \models S \;\Leftrightarrow\; \rho \models P(S),
$
i.e., the current traffic environment and the states of all participating agents satisfy the constraints defined by the predicates in $S$ under the valuation $\rho$.

This evaluation relies on a semantic interface that maps the $AV$ to a runtime monitor based on $S$. 
Specifically, sensor data and other relevant information (e.g., from V2X communication) must be mapped in a type-consistent way to the attributes defined in the $OM$ used by $S$. 
For each object attribute in \autoref{eq:exsv} (e.g., \textit{ego.velocity} in $m/s$), an appropriate input from the $AV$'s available data must be provided according to the corresponding $OM$ (see \autoref{fig:exOM}). 
Only then can the runtime monitor deliver a meaningful verdict on whether $\sigma \models S$.

\subsection{Monitor Verdicts for Explanation Selection}

We first consider a single traffic situation specification $S$.
For this specification, we define a runtime monitor $\mathcal{M}_{S}$ that evaluates whether the concrete traffic situation $\sigma$ where $AV$ is in, corresponds to $S$:

\vspace*{-.4cm}
\begin{equation}\label{eq:msi}
	\mathcal{M}_{S}(\sigma) := 
	\begin{cases} 
		\top & \text{if } \sigma \models S \\ 
		\bot & \text{otherwise} 
	\end{cases}
\end{equation}

Since an explanation model $EM$ describes an $AV$'s behaviour across many traffic situations, it requires multiple explanations $E_i$ with their respective contexts $C_i$, formalised as TSC specifications $S_i$. 
We therefore construct a monitoring system comprising a runtime monitor $\mathcal{M}_{S_i}$ for each $S_i \in \Ss$, where $\Ss$ is the set of all traffic situation specifications defined for $EM$.

We define a function $\mathcal{V}$ that, given the current traffic situation $\sigma$ and a set of explanations $\mathbf{E}$, returns all explanations $E_i \in \mathbf{E}$ whose monitors $\mathcal{M}_{S_i}$ evaluate to $\top$:
\vspace*{-.15cm}
\begin{equation}\label{eq:mem}
	\mathcal{V}(\sigma, \mathbf{E}) :=
	\begin{cases} 
		\{ E_i \in \mathbf{E} \mid \mathcal{M}_{S_i}(\sigma) = \top \} & \text{if } \exists E_i \in \mathbf{E} : \mathcal{M}_{S_i}(\sigma) = \top, \\[1mm]
		\emptyset & \text{otherwise.}
	\end{cases}
\end{equation}
The function thus returns all explanations which correspond to $\sigma$, or $\emptyset$ if none correspond.
Having multiple explanations apply to the same situation can be useful, e.g., during development. To avoid ambiguities for other stakeholders, such as passengers, $\mathbf{E}$ can be required to be \textit{well-formed}, i.e.,
	$\forall E_i, E_j \in \mathbf{E},\, i \neq j:\ \neg \exists \sigma \ (\mathcal{M}_{S_i}(\sigma) = \mathcal{M}_{S_j}(\sigma) = \top)$. For TSC specifications, this can be verified through consistency analysis introduced in~\cite{Becker2024}.

If all runtime monitors $\mathcal{M}_{S_i}(\sigma)$ return $\bot$, two cases may occur:
\begin{itemize}  
	\item[(i)] $AV$ is in a traffic situation $\sigma$ for which no context specification $S_i$ was defined at design time, e.g., an unconsidered or potentially unsafe scenario.
	\item[(ii)] The traffic situation $\sigma$ is covered by $AV$'s explanation model $EM$, but the set of explanations $\mathbf{E}$ is incomplete, missing an $E_i$ for this context.
\end{itemize}
This distinction is important for validation and trust: case (i) can guide extensions of the explanation model, while case (ii) reveals gaps in the existing explanations. 
Additionally, a $\bot$ verdict can trigger runtime data logging, indicate sensor or perception issues, or activate a \textit{safety fallback} in unsafe, uncovered situations.

In conclusion, the goal is that for every concrete traffic situation $\sigma$ that $AV$ might encounter and is described in an explanation model $EM$, there exists an $E_i \in \mathbf{E}$ for each $\sigma \models S_i$. If this is not the case, the $\bot$ verdict acts as a meaningful diagnostic property for incompleteness of explanations, system failures, or domain violations.

With the presented runtime monitoring based on traffic situation specifications $S_i$ linked to explanations $E_i$, we can recognise at runtime whether an $AV$ is in a situation requiring an explanation.
This enables the context-aware presentation of (ante-hoc) explanations, in this work concretely about (un)expectable driving manoeuvres. Recall that combining TSC-based context $C$ formalisation and a dedicated runtime monitoring can be used to trigger any explanations. Regarding the MAB-EX Framework~\cite{Blumreiter2019}, this combination instantiates the \textit{\textbf{M}onitoring} and \textit{\textbf{A}nalyse} phase for triggering context-aware explanations at system runtime. 

\vspace*{-.4cm}
\section{Specification of (Un)expectable Driving Manoeuvres}\label{sec:maneuver-spec}
In the following, we present the specification of driving manoeuvres with Traffic Sequence Charts (TSC), that can be used to specify ante-hoc behaviour explanation of $\AVs$.
We define a \textit{driving manoeuvre} as a purposeful change in the state of attributes of $AV$ over time, taking into account the environment and other traffic participants. We define the annotation \textit{expectable} and \textit{unexpectable} to a driving manoeuvre as follows.
\begin{definition}[(Un)expectable Driving Manoeuvre]
	Given a concrete traffic situation $\sigma$, we call an evolution $\pi : \mathbb{T} \rightarrow ID \rightarrow attr(\mathbf{C}) \rightarrow \mathcal{D}$ starting in $\sigma = \pi(0)$ 
	an expectable (resp. unexpectable) driving manoeuvre if and only if, $AV$ may (resp. may not) continue from $\sigma$ with $\pi$.
\end{definition}
As introduced in \autoref{sec:intro}, the term \textit{(un)expectable} is used to describe manoeuvres that are permitted or not according to an \AV’s explanation model. Unlike strict requirement compliance, this notion captures model-based behaviours while considering domain and AI-based uncertainties.

Referring to our exemplary context $C$ of an $AV$ (see \autoref{ex:context}), a functional description of an expectable and an unexpectable driving manoeuvre can be as follows:

\begin{example}[(Un)expectable Driving Manoeuvres]\label{ex:manoeuvre}
\textit{Given this close distance between $AV$ and the other vehicle ($\leq 25$ meters) and a free adjacent left lane, an expectable manoeuvre is to initiate an overtaking by activating the left indicators and changing lanes. At least in Germany, an unexpectable driving manoeuvre is to initiate an overtaking on the right side (as this violates the traffic rules)}.
\end{example}

Hence, a traffic scenario specification is also a (valid) combination of consecutive traffic situation specifications $S_i$, where each $S_i$ specifies, e.g., a phase of a driving manoeuvre. 

The specification in TSC provides a formalised and context-aware explanation $E := (S, \FE, \FV)$ of (un)expectable driving manoeuvrers, where $S$ is the traffic situation specification, i.e. the context $C$ of $E$, $\FE$ is a non-empty set of specified \textit{expectable} driving manoeuvrers, and $\FV$ is a non-empty set of specified \textit{unexpectable} driving manoeuvrers.

\vspace*{-.3cm}
\subsection{Visual and Formal Specification of Driving Manoeuvres using TSC}

In TSC, \textit{invariant nodes} can be connected consecutively over the same $OM$ and same \textit{symbol dictionary} (for a brief TSC introduction see \Autoref{subsec:tsc-intro}). 
This enables the connection of traffic situation specifications resulting in abstract traffic scenario specifications. Such specifications can formally represent driving manoeuvres, satisfying our goal of specifying (un)expectable behaviour of \AVs.

\begin{figure}[h!]
	\centering
	\begin{subfigure}[b]{0.49\textwidth}
		\centering
		\includegraphics[width=\textwidth]{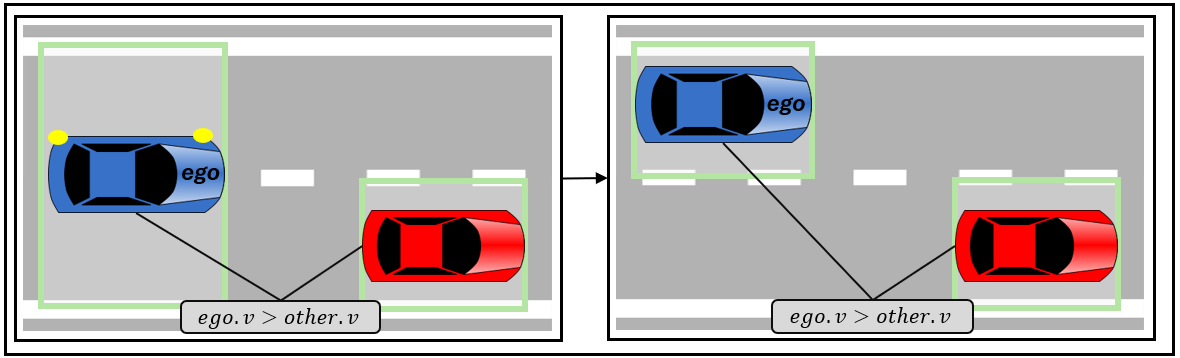}
		\caption{TSC Sequence Chart specifying an exemplary expectable manoeuvrer $F_i \in \FE$ of $AV$ initiating an overtaking of the other vehicle on the adjacent left lane.}
		\label{fig:ex-seq}
	\end{subfigure}
	\hfill
	\begin{subfigure}[b]{0.49\textwidth}
		\centering
		\includegraphics[width=\textwidth]{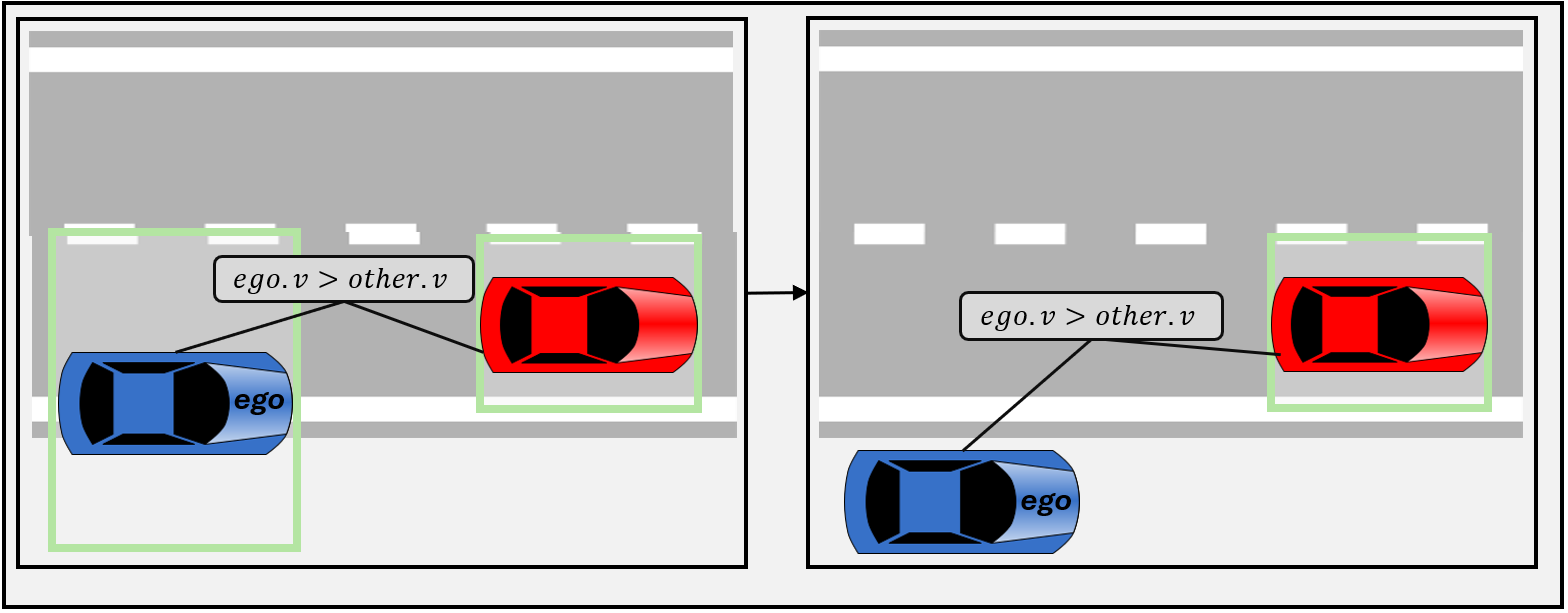}
		\caption{TSC Sequence Chart specifying an exemplary unexpectable manoeuvrer $F_i \in \FV$ of $AV$ initiating overtaking of the other vehicle on the right side.}
		\label{fig:ex-neg}
	\end{subfigure}
	\vspace*{-.2cm}
	\caption{Formalisation of exemplary (un)expectable manoeuvres for our context-aware explanation $E := (S, \FE, \FV)$, specified as abstract traffic scenarios using Traffic Sequence Charts~\cite{atr117}.}
	\label{fig:ex-maneuvers}
	\vspace*{-.4cm}
\end{figure}

By applying available operators of TSC such as sequence, negation, choice or concurrency, invariant nodes can be composed into more complex chart structures (compared to a \textit{Basic Chart}). When composed using the time model $\mathbb{T}$ of TSC, it allows the specification of abstract traffic scenarios, i.e. structures that semantically combine sets of well-typed predicate logic formulae over a given $OM$ using the above operators. Formally, these scenarios can be regarded as expressions using operators over sets $\Phi_i$ of well-typed predicate logic formulae, i.e. expressions of the form $\mathcal{O}(\Phi_1, \Phi_2, \dots, \Phi_n)$, where each $\Phi_i$ is a set of formulas over $OM$ and $\mathcal{O}$ denotes an n-ary operator from the set of available operators in TSC.
For this work, we only consider sequences of invariant nodes. The conditions of each invariant node $N_i$ (such as \autoref{eq:exsv}) are conjugated. The semantics of a sequence chart $SC$ containing two nodes $N_1$ and $N_2$ is defined as  
$
[b,e] \models SC \;:\iff\; \exists t \in [b,e] : [b,t] \models N_1 \wedge [t,e] \models N_2,  
$
where $[b,e]$ is the time interval of evaluation. Further details on composed charts can be found in \cite{atr117,Stemmer2025}.
\autoref{fig:ex-seq} depicts a sequence chart with two invariant nodes.

In the first phase, a car changes to the left lane behind another car. The second phase completes the lane change. This sequence specifies our exemplary expectable manoeuvre. Similarly, \autoref{fig:ex-neg} shows a lane change to the right and its completion, representing our exemplary unexpectable manoeuvre. Both charts use the same $OM$ and \textit{symbol dictionary} as \autoref{fig:ex-tsc}, with added boolean attributes for ego’s indicators and corresponding visual features (yellow blobs).

Predicate logical formulas of the specifications in \autoref{fig:ex-maneuvers} can be derived with the support of algorithms in~\cite{atr117}. For brevity, and because we focus on the visual part of TSC-based manoeuvres as explanation means $M$ at runtime, only the visual representations are shown. To illustrate the formal semantics of one invariant node $N_i$, we provide predicate logical formulas for two context specifications in detail (see \autoref{eq:exsv} and \autoref{fig:eval-phase1-sv}), as needed for constructing runtime monitors (see \autoref{sec:monitoring}). For completeness, the formal semantics for \autoref{fig:ex-maneuvers} can be found in the appendix (see \autoref{tab:phase2-manoeuvres}).

According to our concept, an explanation $E$ is a tuple  
$
E := (S_i, \FE, \FV),  
$
where $S_i \in \Ss$ specifies the context $C$ and explanandum $X$, $\FE$ is the set of expectable manoeuvres, and $\FV$ is the set of unexpectable manoeuvres, defined by the AV’s explanation model $EM$. By specifying our exemplary context $C$ (see \autoref{fig:ex-tsc}) and the (un)expectable manoeuvres (see \autoref{fig:ex-maneuvers}), we have all components for an explanation $E$ using the visual yet formal TSC language.
Given that $\FE$ and $\FV$ are correctly associated with $S_i \in \Ss$ in the explanation model $EM$ of $AV$, the explanation $E = (S_i, \FE, \FV)$ becomes context-aware. Using the same $OM$ and symbol dictionary, the abstract traffic situation specification $S_i$ of the explanation context $C$ determines \textit{when} an explanation is required, while the associated manoeuvre sets $\FE$ and $\FV$ define \textit{what} is to be communicated. In \Autoref{sec:experiment}, we show a possible way of presenting such an explanation $E$.

In contrast to common XAI approaches, our method enables consistent alignment of context and explanation content already during design. 
This may benefit a variety of stakeholders: Testing engineers can validate that decision-making in specific traffic situations aligns with the formalised (un)expectable manoeuvres. Passengers or safety assessors may be informed why a specific manoeuvre is being (or not being) executed, increasing transparency and trust.

By combining explanation context and content specification using TSC with a dedicated runtime monitoring, we operationalise the loop between formal specification, runtime monitoring, and context-aware explanations, as conceptualised by the MAB-EX framework~\cite{Blumreiter2019}. Hence, this approach supports the development of trustworthy autonomous driving systems.

\newcommand{\FEi}[1]{\mathcal{F}_{E_{#1}}}
\newcommand{\FVi}[1]{\mathcal{F}_{V_{#1}}}

\section{Application of Context-aware, Ante-hoc TSC-Explanations}
\label{sec:experiment}
In this section, we demonstrate the applicability of our approach for presenting context-aware, ante-hoc explanations of (un)expectable driving manoeuvres and validate its runtime capability.
Concretely, we specify multiple explanations $E_i$ using TSC.
Each explanation comprises an explanation context $C_i$, formalised as an abstract traffic situation $S_i \in \Ss$, together with the corresponding (un)expectable driving manoeuvres ($\FE$ and $\FV$).
We then construct runtime monitors $\mathcal{M}_{S_i}(\sigma)$ for each context $C_i$ and embed them into a simulation.
Finally, we provide a visual yet formal, ante-hoc presentation of the (un)expectable manoeuvres.
Our approach is demonstrated in a simulated overtaking scenario of an $\AV$.

To this end, \Autoref{subsec:setup} describes the experimental setup.
In \Autoref{subsec:eval-expl}, we present the specified explanations $E_i$, each linking an abstract traffic situation $S_i$ to individual phases of an overtaking manoeuvre: approaching, closing the gap, changing lanes, and overtaking.
\Autoref{subsec:eval-mon} then shows the runtime monitoring results and the ante-hoc explanations of (un)expectable manoeuvres.
Finally, \Autoref{subsec:eval-disc} discusses our method.
\vspace*{-.4cm}
\subsection{Experimental Setup}\label{subsec:setup}
The hardware specification is reported to ensure reproducibility, comparability, and to provide a context for later runtime results. The hardware consists of $192\,\mathrm{GB}$ of RAM, an Intel\textsuperscript{\textregistered} Xeon\textsuperscript{\textregistered} Silver~4215R CPU (16~cores @ $3.20\,\mathrm{GHz}$), and an NVIDIA\textsuperscript{\textregistered} Quadro RTX~6000 GPU.
CARLA~\cite{Dosovitskiy2017} (version 0.9.15) was used as simulation environment for simulating the overtaking scenario.  
In the simulated traffic scenario, an $AV$ (car, red) drives towards a slower-moving vehicle (van, grey) in the right-hand lane of a two-lane carriageway. At a certain distance, the $AV$ initiates an overtaking manoeuvre, passes the vehicle to be overtaken and initiates the final lane change.

The runtime monitors were implemented in Python and connected to the CARLA simulation via the ROS2–CARLA Bridge \cite{carla_ros_bridge_github}. Data were continuously transmitted at a frequency of 20 Hz during simulation runtime.
We do not use CARLA's Python API directly, but rather ROS 2 as the communication protocol for $AV$ data, as many real-world systems utilise ROS 2 for data transmission. This has enabled TSC runtime monitoring to be successfully deployed in a maritime context~\cite{Austel2025}, directly from the simulation onto a real vessel.

Since our runtime monitors are constructed from abstract traffic situation specifications $S_i \in \Ss$ defined using TSCs, each monitor relies on the corresponding $OM$ of the TSC specification. The object type instances and their respective attributes defined in the $OM$ were mapped via an interface to suitable, available data provided by the CARLA simulation. Recall that only through appropriate mapping of objects and their attributes to suitable system data and signals can runtime monitoring deliver meaningful verdicts.  
Once a runtime monitor yields $\mathcal{M}_{S_i}(\sigma) = \top$ (see \autoref{eq:msi}), indicating that the $AV$ is in the corresponding traffic situation $\sigma$, the related explanation $E_i$ for expectable ($\FE$) and unexpectable ($\FV$) manoeuvres is presented. 
The function $\mathcal{V}(\sigma, \mathbf{E})$ (see \autoref{eq:mem}) that returns context-aware explanations for the concrete traffic situation $\sigma$ of the $AV$ was implemented in Python. It selects and triggers the presentation of corresponding explanations $E_i$.
A video of the case study is provided in \cite{ApplicationVideo}.

\vspace*{-.3cm}
\subsection{Defined Explanations}\label{subsec:eval-expl}
To demonstrate and validate our context-aware, ante-hoc explanation method, we decomposed the simulated overtaking into distinct phases. Each phase describes a context and its corresponding (un)expectable manoeuvres, which are visually and formally specified using TSC. Hence, the specification of each phase serves as our explanation in a tuple $E = (S_i, \FE, \FV)$.
The following distinct phases of the considered overtaking are incorporating our example introduced earlier in \Autoref{sec:tsc-explanations} and \Autoref{sec:maneuver-spec} (here \textit{Phase 2 - Closing Gap}):

\textbf{{\small Phase 1 - Approaching.}} \textit{The $AV$ is driving towards a slower driving vehicle in its lane. There is no other vehicle in the adjacent left lane of ego. The distance between ego and the other vehicle is greater than 25 metres. Given the moderate distance in this traffic situation, an expectable manoeuvres are (i) to initiate an overtaking by activating the left indicators and changing lanes, or (ii) to adapt to the speed of the leading vehicle and stay behind at a safe distance. An unexpectable driving manoeuvre is to initiate an overtaking on the right side (e.g., because of violating the traffic rules of German traffic law).}

\begin{figure}[htbp]
	\centering
	\begin{minipage}[c]{0.43\textwidth} 
		\centering
		\includegraphics[width=0.8\linewidth]{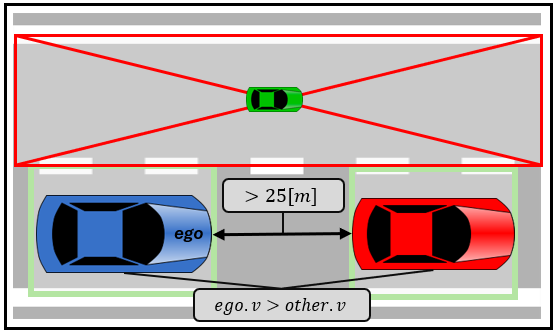}
		\subcaption{Specified Basic Chart}
	\end{minipage}%
	\hfill
	\begin{minipage}[c]{0.56\textwidth} 
		\begin{subequations}\label{eq:phase1}
			\begin{align*}
				S_1 = &ego.pos.y > lane.yR \land ego.pos.y < lane.yM \\
				\land\text{ }&other.pos.y > lane.yR \land other.pos.y < lane.yM \\
				\land\text{ }&ego.pos.x + 25 > other.pos.x \\
				\land\text{ }&ego.v > other.v\\
				\land\text{ }&ego.inRange(left) = \emptyset
			\end{align*}
		\end{subequations}
		\vspace*{-.8cm}
		\subcaption{Corresponding predicate logical formula}
	\end{minipage}
	\caption{TSC specification of described traffic situation of $AV$ in \textit{Phase 1 - Approaching}, using the object model $OM$ \autoref{fig:exOM} and symbol dictionary  \autoref{fig:exSD}}
	\label{fig:eval-phase1-sv}
	\vspace*{-.1cm}
\end{figure}

The specification of the abstract traffic situation $S_i$ for \textit{Phase 1} as TSC, as well as the derived predicate logical formula, is depicted in \autoref{fig:eval-phase1-sv}.  

\begin{wrapfigure}[22]{r}{0.45\textwidth}
	\centering
	\vspace*{-.5em}
	\includegraphics[scale=0.23]{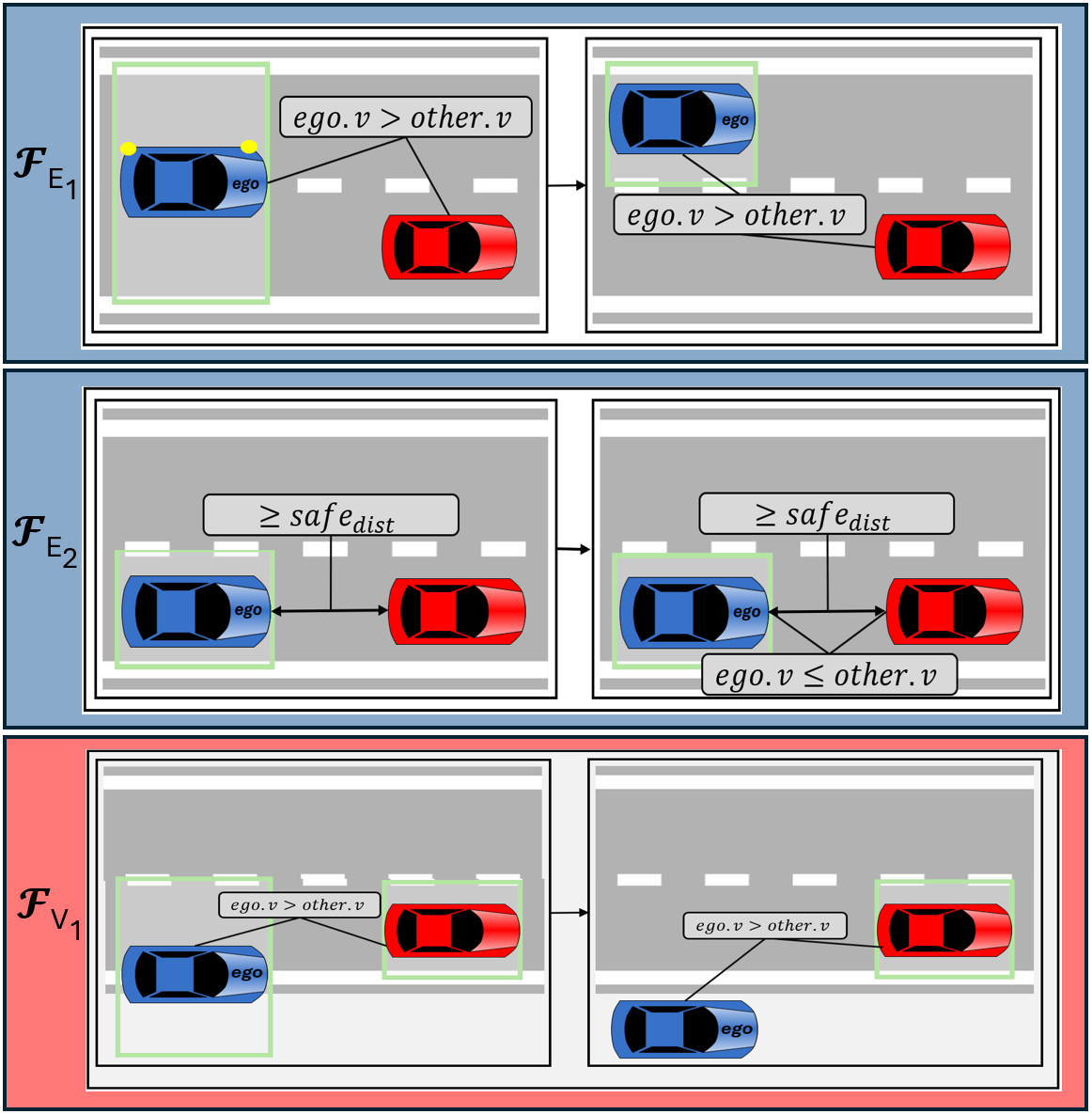}
	\caption{Representation of (un)expectable driving manoeuvres ($\FEi{1},\FEi{2},\FVi{1}$) of an explanation $E_i \in \mathbf{E}$ for \textit{Phase 1 - Approaching} (see \autoref{subsec:eval-expl}), specified as abstract traffic scenarios using TSC~\cite{atr117}.}
	\label{fig:eval-phase1-man}
\end{wrapfigure}

The visual specifications of described (un)expectable driving manoeuvres for \textit{Phase 1} are depicted in \autoref{fig:eval-phase1-man}.Readers interested in the formal semantics of the visual specifications are referred to the appendix (see\autoref{tab:phase1-manoeuvres}).

For presentation at runtime, we embedded the specifications of driving manoeuvres in a colored frame. This is just one possible way of presenting our explanations. 

We chose a blue frame to communicate the case of more than one expectable driving manoeuvre. If there is only one expectable driving manoeuvre, we choose a green colored frame. We choose a red frame for every unexpectable driving manoeuvre.

\textbf{{\small Phase 2 - Closing Gap.}}\textit{ $AV$ is following a slower driving vehicle (slower than speed limit) at a distance $\leq 25$ meters. Given this close distance between $AV$ and the other vehicle ($\leq 25$ meters) and a free adjacent left lane, an expectable manoeuvre is to initiate an overtaking by activating the left indicators and changing lanes. An unexpectable driving manoeuvre is to initiate an overtaking on the right side (e.g., because of violating the traffic rules in the German StVO).}

The specified explanation $E_2$ of \textit{Phase 2} using TSC has already been conducted in \Autoref{sec:tsc-explanations}. The colour-coded version for the presentation in simulation of the (un)expectable driving manoeuvres is depicted in \autoref{tab:overtaking_explanations} Readers interested in the formal semantics of the visual specifications are referred to the appendix (see \autoref{tab:phase2-manoeuvres}).

\textbf{{\small Phase 3 - Lane Change.}}\textit{ $AV$ is initiating an overtaking of a slower driving vehicle (slower than speed limit) by changing lanes with active left indicators. An expectable driving manoeuvre is to reach the adjacent left lane and begin passing the slower-moving vehicle. An unexpectable driving manoeuvre, given to the free target lane, is to abort the lane change, driving slower, and then driving behind the other vehicle at a safe distance.} 

The explanation context of \textit{Phase 3}, specified as abstract traffic situation $S_3$, along with the colour-coded explanation $E_3$ showing the (un)expectable manoeuvres, is depicted in \autoref{tab:overtaking_explanations}. Readers interested in the formal semantics of the visual specifications are referred to the appendix (see \autoref{tab:phase3-manoeuvres}).

\textbf{{\small Phase 4 - Overtaking.}} \textit{$AV$ is driving on the left lane of a two-lane carriageway and is about to overtake a slower driving vehicle (slower than the speed limit) on the right lane. An expectable driving manoeuvre is to initiate a lane change to the right lane and in front of the vehicle to be overtaken. An unexpectable driving manoeuvre is to drive slower than the other vehicle and to abort the overtaking by a lane change to the right behind the other vehicle.}

The explanation context of \textit{Phase 4}, specified as abstract traffic situation $S_4$, along with the colour-coded explanation $E_4$ showing the (un)expectable manoeuvres, is depicted in \autoref{tab:overtaking_explanations}. Readers interested in the formal semantics of the visual specifications are referred to the appendix (see \autoref{tab:phase4-manoeuvres}).

\begin{table}[h!]
	\vspace{-.3cm}
	\centering
	\caption{Overview of context specifications $C_i$ and corresponding (un)expectable driving manoeuvres ($E_i \in \mathbf{E}$) for the consecutive phases 2-4 of the overtaking scenario. Each context is specified as a TSC-based traffic situation specification. The corresponding explanation is specified as an abstract traffic scenario according to the object model $OM$ (\autoref{fig:exOM}) and symbol dictionary (\autoref{fig:exSD}).}
	\renewcommand{\arraystretch}{1.2}
	\setlength{\tabcolsep}{6pt}
	\resizebox{\textwidth}{!}{%
		\begin{tabular}{
				@{}  
				>{\centering\arraybackslash}m{2.1cm}
				>{\centering\arraybackslash}m{3.7cm}
				>{\centering\arraybackslash}m{6.5cm}
				@{}  
			}
			\toprule
			{\footnotesize \textbf{Phase}} & {\footnotesize \textbf{Context}} & {\footnotesize \textbf{(Un)expectable Manoeuvrers}} \\
			\midrule
			{\footnotesize Phase 2 – Closing Gap} &
			\includegraphics[scale=0.16]{figures/exampleSV.png} &
			\includegraphics[scale=0.2]{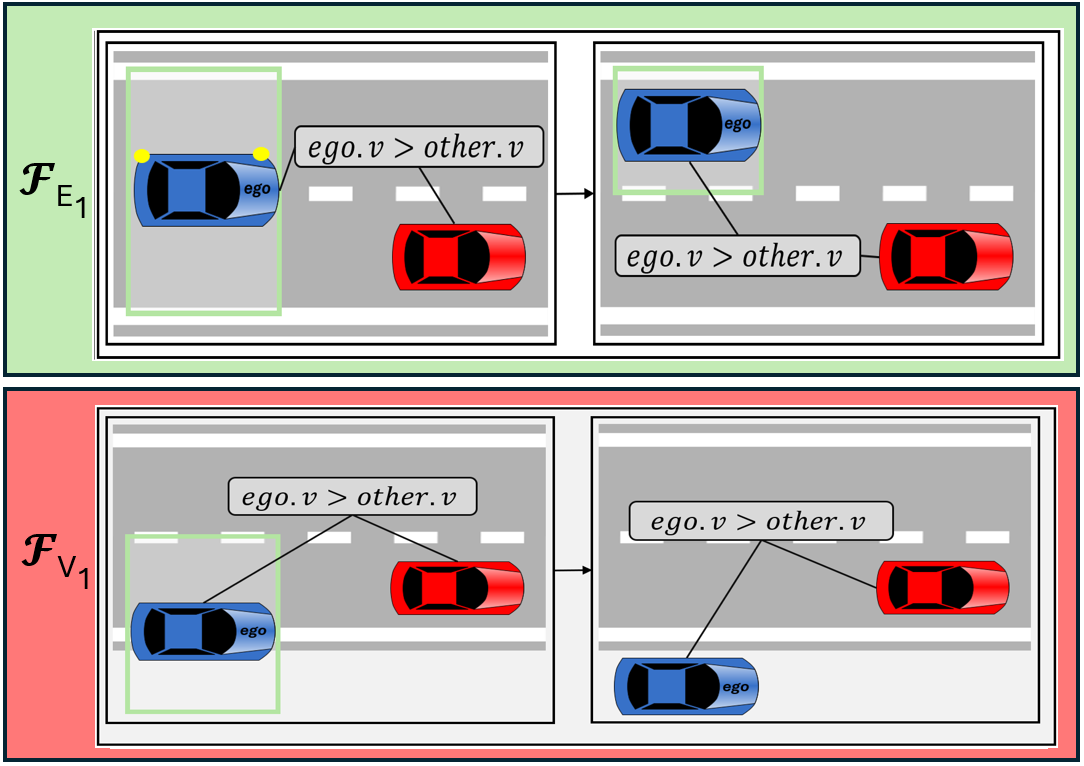} \\ \midrule[.5pt]
			{\footnotesize Phase 3 – Lane Change} &
			\includegraphics[width=0.2\textwidth]{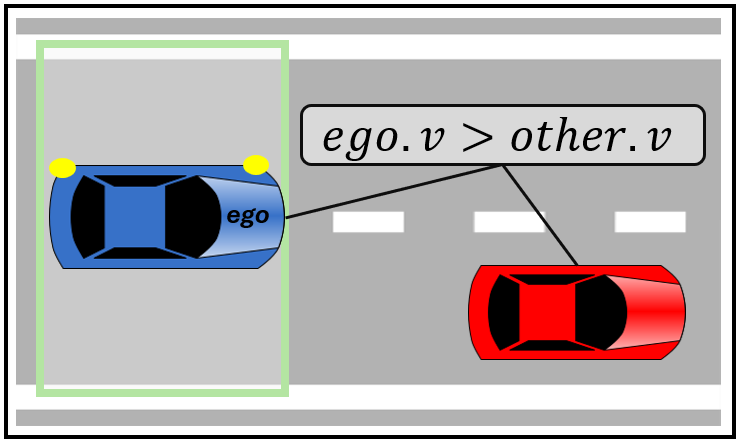} &
			\includegraphics[scale=0.21]{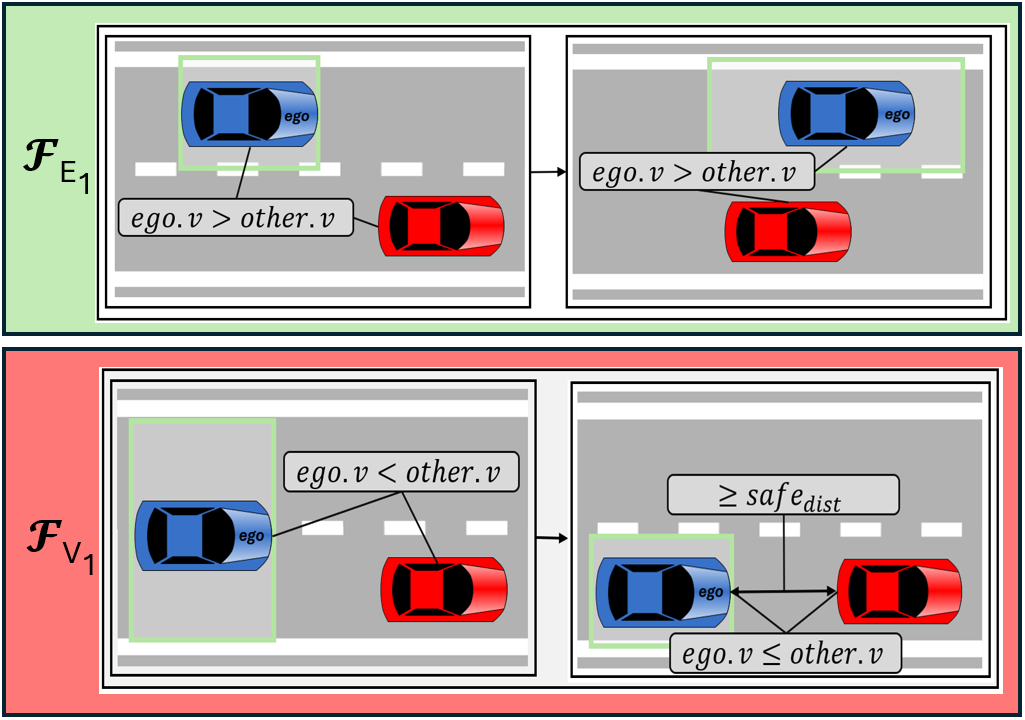} \\ \midrule[.5pt]
			{\footnotesize Phase 4 – Overtaking} &
			\includegraphics[width=0.2\textwidth]{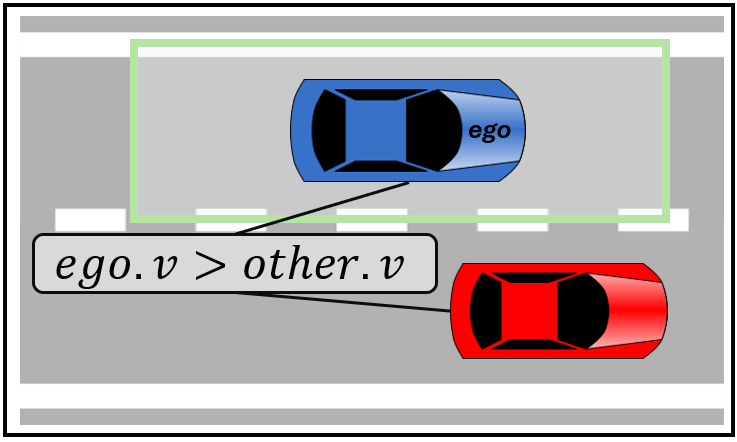} &
			\includegraphics[scale=0.2]{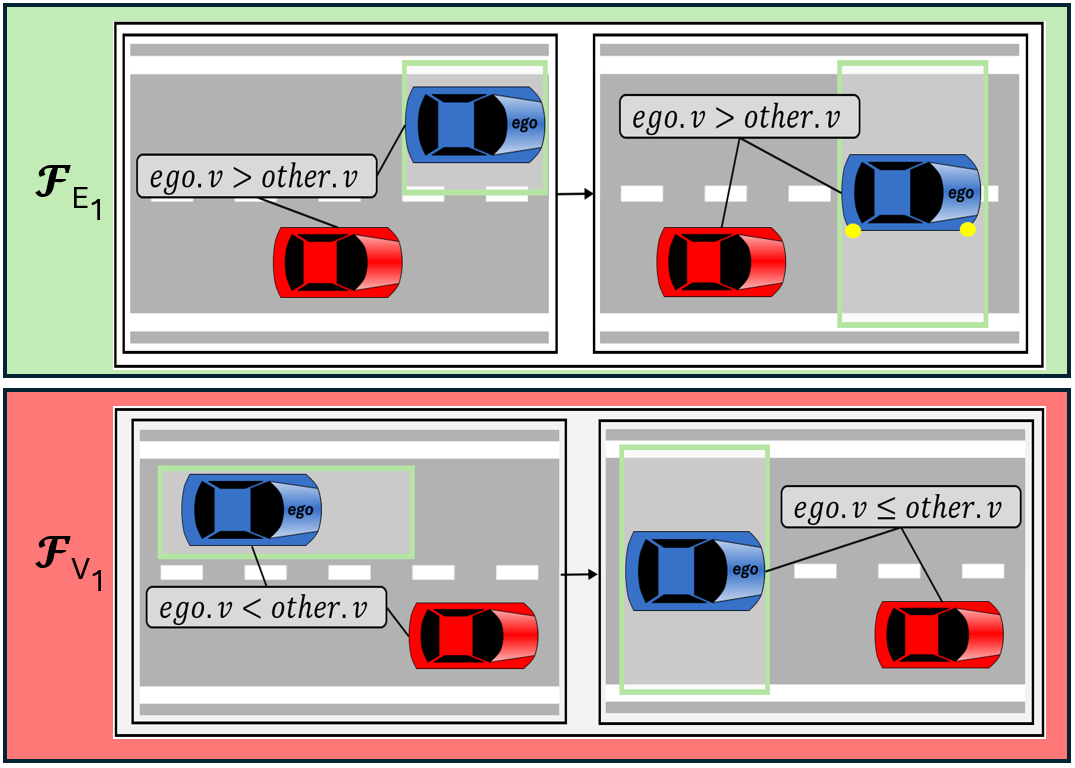} \\
			\bottomrule
		\end{tabular}%
	}
	\label{tab:overtaking_explanations}
	\vspace*{-.5cm}
\end{table}

\subsection{Runtime Detection and Explanation Timing}\label{subsec:eval-mon}
For runtime monitoring in the simulation, the class instances and attributes from the defined $OM$ (\autoref{fig:exOM}) were directly mapped to available simulation signals. A reliable interface between vehicle sensor data and runtime monitors is essential: We converted measurements (e.g., bounding boxes, speeds) into typed $OM$ attributes (position $\in \mathbb{R}^2$, velocity $\in \mathbb{R}$, lane ID, etc.). Corresponding to the four overtaking phases, we constructed four runtime monitors, each evaluating a predicate formula $P(S_i)$ derived from TSC specifications of contexts $C_i$. When $\varrho \models P(S_i)$, the associated explanation $E_i$ is selected and presented ante-hoc, i.e., before execution. For each $E_i \in \mathbf{E}$, the corresponding runtime monitor $\mathcal{M}_{S_i}(\sigma)$ is embedded in the simulation, as detailed in \Autoref{subsec:setup}.

\begin{figure}[h]
	\centering
	\includegraphics[scale=0.34]{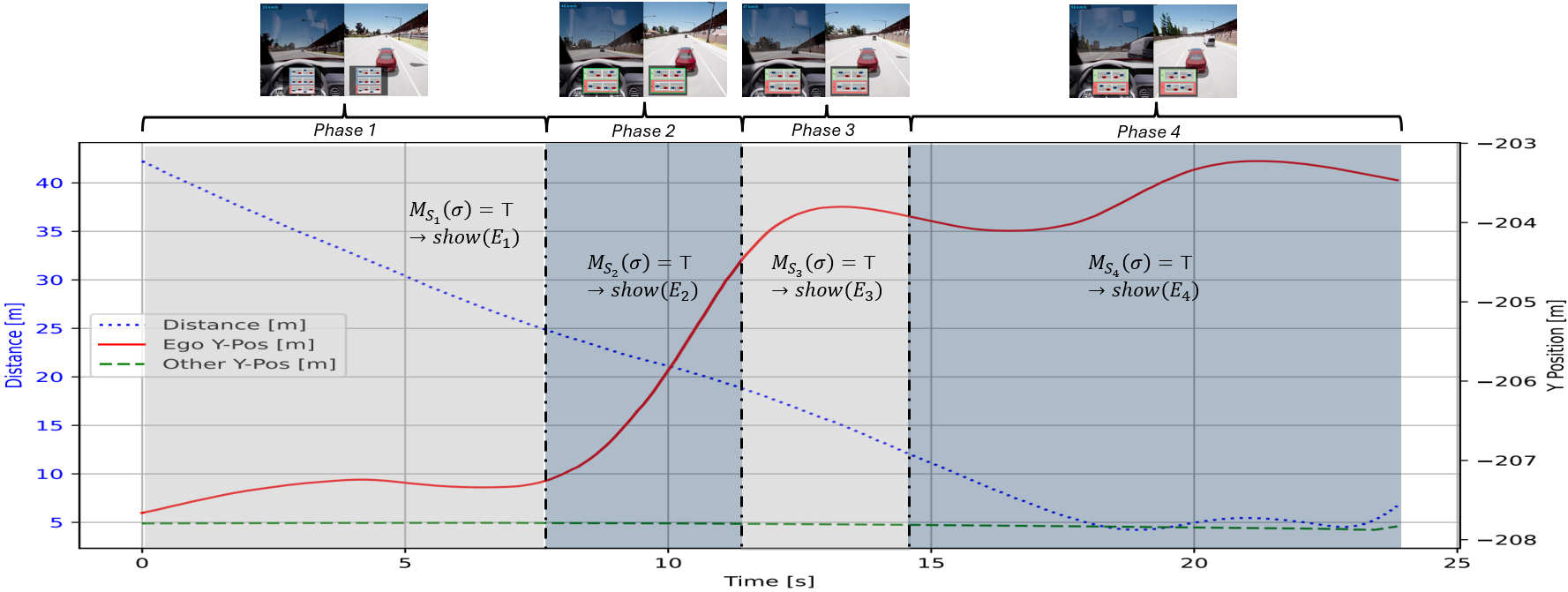}
	\caption{Important attributes (distance between $AV$ (ego) and the van (other), blue graph; y-position of $AV$ (ego), red graph; and y-position of the van (other), green graph) of the simulated overtaking scenario presenting the decomposed phases with images of the simulation above and length, marked with brackets over the time dimension. The blue-ish and grey-ish frames describe the time at which the respective monitors $\mathcal{M}_{S_i}(\sigma)$ recognised a corresponding concrete traffic situation $\sigma$, as well as which explanation $E_i$ was subsequently presented. The dashed vertical lines describe the time at which another runtime monitor recognises the corresponding traffic situation.}
	\label{fig:eval-rm}
\end{figure}

\autoref{fig:eval-rm} visualises the evolution of relevant attributes in the simulated overtaking over time. It also outlines at what point in time which monitor $\mathcal{M}_{S_i}(\sigma)$ recognises a corresponding concrete traffic situation $\sigma$, i.e. $\mathcal{M}_{S_i}(\sigma) = \top$ (blue-ish/grey-ish frames). The indexing of the monitors corresponds to the phases described, i.e. $\mathcal{M}_{S_1}(\sigma)$ is the runtime monitor constructed from the abstract traffic situation specification $S_1$ and \textit{Phase 1 - Approaching} (see \autoref{fig:eval-phase1-sv}).

The runtime monitors $\mathcal{M}_{S_i}(\sigma)$ were able to recognise the corresponding explanation contexts, i.e., evaluate $\sigma \models S_i$, in under 1 $\mathrm{ms}$ after the $\AV$ entered the respective traffic situation $\sigma$ in the simulation.
This shows that TSC-based runtime monitoring can reliably detect $\sigma \models S_i$. It also enables the context-aware presentation of explanations, effectively operationalising the \textit{Monitoring} and \textit{Analysis} phases of the MAB-EX framework~\cite{Blumreiter2019}.

\autoref{fig:eval-rm} also shows which explanation $E_i$ was presented and for how long. Since each explanation $E_i$ for an abstract traffic situation $S_i$ is correctly associated with its (un)expectable driving manoeuvres, it demonstrates that the context-aware explanation $E_i$ was always presented when the $AV$ was in a corresponding concrete traffic situation $\sigma \models S_i$. Hence, this demonstrates the operationalisation of phases \textit{Build} and \textit{EXplain} of the MAB-EX framework~\cite{Blumreiter2019}.
Since the respective explanation $E_i$ was presented immediately after $\mathcal{M}_{S_i}(\sigma) = \top$, and every respective traffic situation is present for more than 2 seconds in the simulation, our explanation of (un)expectable driving manoeuvres is ante-hoc.   
\vspace*{-.4cm}
\subsection{Discussion}\label{subsec:eval-disc}
Our demonstration validates in a simulated overtaking that our approach is able to provide (i) context-aware, ante-hoc triggering of explanations, and (ii) visually yet formally specified (un)expectable driving manoeuvres before an $AV$ is manoeuvring.
The explanations were specified in a visual yet formal manner using TSC, designed to be context-aware, and combined with a dedicated runtime monitoring.

By implementing our approach in the simulation environment {\small CARLA}, we have demonstrated its capabilities. Explanations can be triggered ante-hoc (before a manoeuvre is executed), based on the concrete traffic situation $\sigma$ in which the $AV$ is currently in and which corresponds to a formal explanation context specified as an abstract traffic situation $S_i$.
Regarding the scalability of our approach,  there is tool support for the creation of TSC specifications and the generation of predicate logic formulas~\cite{Borchers2025}. Executing several runtime
monitors in simulation is not a limiting factor.
The demonstration was restricted to a controlled and simulated overtaking and did not include testing in more complex environments. The simulation setup also relied on idealised sensor data and environmental models, without accounting for factors such as sensor noise.
Further, to investigate whether a TSC-based explanation is indeed able to fulfil the goal of increasing trust, empirical studies need to be conducted with stakeholder groups. Hence, for now, we can not provide any empirical statement if an explanation $E$ specified by TSC is \enquote{\textit{\ldots a piece of information (or evidence) that makes the explanandum $X$ understandable by $G$ with respect to the goal $\theta$}.}\cite{Schwammberger2024xAI}.
However, we believe that the combination of visual syntax and rigorous formal semantics based on predicate logic allows the same TSC explanation to address different stakeholders. For example, the visual representation of $E$ could address passengers of $AV$, while the formal semantics could address developers or test engineers.

\vspace*{-.5cm}
\section{Conclusion and Future Work}
\label{sec:conclusion}
This work aims to increase trust in autonomous vehicles ($\AVs$) by enabling context-aware, ante-hoc explanations at system runtime to support the bridging of 
correct (by rigorous formal semantics) and good (by both visual and formal means) explanations.
Within the field of Explainability Engineering, we present a formalisation of explanation contexts, i.e., a traffic situation of an $\AV$ where an explanation is meant to be presented, using Traffic Sequence Charts (TSC). In combination with a dedicated TSC runtime monitoring, we can observe the current traffic situation of $\AVs$, evaluate if it corresponds to the formalised explanation context, and trigger the presentation of explanations. Additionally, we show how TSC can formalise (un)expectable driving manoeuvres in a context-aware, visual yet formal way, enabling the same explanation to be used for different stakeholders. In a simulated overtaking scenario, we demonstrate and validate our contribution of context-aware explanations of (un)expectable driving manoeuvres, and show that we were able to present ante-hoc explanations, i.e., before an $\AV$ is manoeuvring, at system runtime.
This work aligns with the MAB-EX framework~\cite{Blumreiter2019} and is the first concrete operationalisation of key phases that demonstrates the runtime applicability for context-aware, ante-hoc explanations.

To enhance the applicability and impact of our approach, several extensions are envisaged.
Future work may expand the approach to various and more complex driving scenarios, incorporating uncertainty-aware runtime monitoring techniques (cf.~\cite{Finkbeiner2022}).
This may include empirical studies with diverse stakeholders, designing multimodal explanations (e.g., textual annotations or audio cues), and evaluating feasibility on automotive-grade hardware. Furthermore, we see a potential in combining TSC specifications of explanation contexts with the explanation history extension of the MAB-EX framework~\cite{Schwammberger2024xAI}, as such abstract specifications can generalise over several explanation contexts. 
\bibliographystyle{eptcs}
\bibliography{main}

\begin{thebibliography}{10}
\providecommand{\bibitemdeclare}[2]{}
\providecommand{\surnamestart}{}
\providecommand{\surnameend}{}
\providecommand{\urlprefix}{Available at }
\providecommand{\url}[1]{\texttt{#1}}
\providecommand{\href}[2]{\texttt{#2}}
\providecommand{\urlalt}[2]{\href{#1}{#2}}
\providecommand{\doi}[1]{doi:\urlalt{https://doi.org/#1}{#1}}
\providecommand{\eprint}[1]{arXiv:\urlalt{https://arxiv.org/abs/#1}{#1}}
\providecommand{\bibinfo}[2]{#2}

\bibitemdeclare{misc}{euaiact}
\bibitem{euaiact}
 (\bibinfo{year}{2024}): \emph{\bibinfo{title}{Regulation (EU) 2024/1689 of the
  European Parliament and of the Council of 13 June 2024 laying down harmonised
  rules on artificial intelligence (Artificial Intelligence Act)}}.
\newblock \urlprefix\url{https://eur-lex.europa.eu/eli/reg/2024/1689/oj/eng}.
\newblock \bibinfo{note}{Official Journal of the European Union, L 1689}.

\bibitemdeclare{article}{Alicioglu2022}
\bibitem{Alicioglu2022}
\bibinfo{author}{Gulsum \surnamestart Alicioglu\surnameend} \&
  \bibinfo{author}{Bo~\surnamestart Sun\surnameend} (\bibinfo{year}{2022}):
  \emph{\bibinfo{title}{A survey of visual analytics for Explainable Artificial
  Intelligence methods}}.
\newblock {\slshape \bibinfo{journal}{Computers and Graphics}}
  \bibinfo{volume}{102}, pp. \bibinfo{pages}{502--520},
  \doi{10.1016/j.cag.2021.09.002}.

\bibitemdeclare{article}{Alqahtani2025}
\bibitem{Alqahtani2025}
\bibinfo{author}{Thaar \surnamestart Alqahtani\surnameend}
  (\bibinfo{year}{2025}): \emph{\bibinfo{title}{Recent Trends in the Public
  Acceptance of Autonomous Vehicles: A Review}}.
\newblock {\slshape \bibinfo{journal}{Vehicles}}
  \bibinfo{volume}{7}(\bibinfo{number}{2}), \doi{10.3390/vehicles7020045}.

\bibitemdeclare{inproceedings}{Arechiga2019}
\bibitem{Arechiga2019}
\bibinfo{author}{Nikos \surnamestart Arechiga\surnameend}
  (\bibinfo{year}{2019}): \emph{\bibinfo{title}{Specifying Safety of Autonomous
  Vehicles in Signal Temporal Logic}}.
\newblock In: {\slshape \bibinfo{booktitle}{2019 IEEE Intelligent Vehicles
  Symposium (IV)}}, pp. \bibinfo{pages}{58--63},
  \doi{10.1109/IVS.2019.8813875}.

\bibitemdeclare{article}{Atf2025}
\bibitem{Atf2025}
\bibinfo{author}{Zahra \surnamestart Atf\surnameend} \&
  \bibinfo{author}{Peter~R. \surnamestart Lewis\surnameend}
  (\bibinfo{year}{2025}): \emph{\bibinfo{title}{Is Trust Correlated With
  Explainability in AI? A Meta-Analysis}}.
\newblock {\slshape \bibinfo{journal}{IEEE Transactions on Technology and
  Society}}, pp. \bibinfo{pages}{1--8}, \doi{10.1109/TTS.2025.3558448}.

\bibitemdeclare{inproceedings}{Austel2025}
\bibitem{Austel2025}
\bibinfo{author}{Anna \surnamestart Austel\surnameend}, \bibinfo{author}{Lukas
  \surnamestart Panneke\surnameend}, \bibinfo{author}{Janusz~Andrzej
  \surnamestart Piotrowski\surnameend}, \bibinfo{author}{Nina \surnamestart
  Wetzig\surnameend}, \bibinfo{author}{Matthias \surnamestart
  Steidel\surnameend} \& \bibinfo{author}{Bernd \surnamestart
  Westphal\surnameend} (\bibinfo{year}{2025}): \emph{\bibinfo{title}{Using
  Monitoring of Maritime Traffic Scenarios in the Validation of Maritime
  Systems}}.
\newblock In: {\slshape \bibinfo{booktitle}{2025 Symposium on Maritime
  Informatics and Robotics (MARIS)}}, \doi{10.1109/MARIS64137.2025.11139541}.
\newblock \urlprefix\url{https://elib.dlr.de/215965/}.

\bibitemdeclare{inproceedings}{Bairy2023}
\bibitem{Bairy2023}
\bibinfo{author}{A.~\surnamestart Bairy\surnameend} \&
  \bibinfo{author}{M.~\surnamestart Fr{\"a}nzle\surnameend}
  (\bibinfo{year}{2023}): \emph{\bibinfo{title}{Optimal Explanation Generation
  using Attention Distribution Model}}.
\newblock In \bibinfo{editor}{Tareq \surnamestart Ahram\surnameend} \&
  \bibinfo{editor}{Redha \surnamestart Taiar\surnameend}, editors: {\slshape
  \bibinfo{booktitle}{Human Interaction and Emerging Technologies (IHIET-AI
  2023): Artificial Intelligence and Future Applications}}, {\slshape
  \bibinfo{series}{AHFE Open Access}}~\bibinfo{volume}{70},
  \bibinfo{publisher}{AHFE International}, \bibinfo{address}{USA},
  \doi{10.54941/ahfe1002928}.

\bibitemdeclare{incollection}{Bartocci2018IntroRV}
\bibitem{Bartocci2018IntroRV}
\bibinfo{author}{Ezio \surnamestart Bartocci\surnameend},
  \bibinfo{author}{Yli\`es \surnamestart Falcone\surnameend},
  \bibinfo{author}{Adrian \surnamestart Francalanza\surnameend} \&
  \bibinfo{author}{Giles \surnamestart Reger\surnameend}
  (\bibinfo{year}{2018}): \emph{\bibinfo{title}{Introduction to Runtime
  Verification}}.
\newblock In: {\slshape \bibinfo{booktitle}{Lectures on Runtime Verification.
  Introductory and Advanced Topics}}, {\slshape \bibinfo{series}{Lecture Notes
  in Computer Science}} \bibinfo{volume}{10457}, \bibinfo{publisher}{Springer},
  pp. \bibinfo{pages}{1--33}, \doi{10.1007/978-3-319-75632-5\_1}.

\bibitemdeclare{article}{Bauer2011}
\bibitem{Bauer2011}
\bibinfo{author}{Andreas \surnamestart Bauer\surnameend},
  \bibinfo{author}{Martin \surnamestart Leucker\surnameend} \&
  \bibinfo{author}{Christian \surnamestart Schallhart\surnameend}
  (\bibinfo{year}{2011}): \emph{\bibinfo{title}{Runtime Verification for LTL
  and TLTL}}.
\newblock {\slshape \bibinfo{journal}{ACM Trans. Softw. Eng. Methodol.}}
  \bibinfo{volume}{20}(\bibinfo{number}{4}), \doi{10.1145/2000799.2000800}.

\bibitemdeclare{inproceedings}{Becker2024}
\bibitem{Becker2024}
\bibinfo{author}{Jan~Steffen \surnamestart Becker\surnameend}
  (\bibinfo{year}{2024}): \emph{\bibinfo{title}{A Consistency Analysis Method
  for Traffic Sequence Charts}}.
\newblock In: {\slshape \bibinfo{booktitle}{VEHITS24 Doctoral Consortium}},
  \doi{10.48550/arXiv.2409.03774}.
\newblock \urlprefix\url{https://elib.dlr.de/204347/}.

\bibitemdeclare{inproceedings}{Blumreiter2019}
\bibitem{Blumreiter2019}
\bibinfo{author}{Mathias \surnamestart Blumreiter\surnameend},
  \bibinfo{author}{Joel \surnamestart Greenyer\surnameend},
  \bibinfo{author}{Francisco~Javier \surnamestart Chiyah~Garcia\surnameend},
  \bibinfo{author}{Verena \surnamestart Klös\surnameend},
  \bibinfo{author}{Maike \surnamestart Schwammberger\surnameend},
  \bibinfo{author}{Christoph \surnamestart Sommer\surnameend},
  \bibinfo{author}{Andreas \surnamestart Vogelsang\surnameend} \&
  \bibinfo{author}{Andreas \surnamestart Wortmann\surnameend}
  (\bibinfo{year}{2019}): \emph{\bibinfo{title}{Towards Self-Explainable
  Cyber-Physical Systems}}.
\newblock In: {\slshape \bibinfo{booktitle}{2019 ACM/IEEE 22nd International
  Conference on Model Driven Engineering Languages and Systems Companion
  (MODELS-C)}}, pp. \bibinfo{pages}{543--548},
  \doi{10.1109/MODELS-C.2019.00084}.

\bibitemdeclare{misc}{bohlender2019characterization}
\bibitem{bohlender2019characterization}
\bibinfo{author}{Dimitri \surnamestart Bohlender\surnameend} \&
  \bibinfo{author}{Maximilian~A. \surnamestart Köhl\surnameend}
  (\bibinfo{year}{2019}): \emph{\bibinfo{title}{Towards a Characterization of
  Explainable Systems}}.
\newblock \eprint{1902.03096}.

\bibitemdeclare{article}{Borchers2025}
\bibitem{Borchers2025}
\bibinfo{author}{Philipp \surnamestart Borchers\surnameend},
  \bibinfo{author}{Tjark \surnamestart Koopmann\surnameend},
  \bibinfo{author}{Lukas \surnamestart Westhofen\surnameend},
  \bibinfo{author}{Jan~Steffen \surnamestart Becker\surnameend},
  \bibinfo{author}{Lina \surnamestart Putze\surnameend},
  \bibinfo{author}{Dominik \surnamestart Grundt\surnameend},
  \bibinfo{author}{Thies \surnamestart {de Graaff}\surnameend},
  \bibinfo{author}{Vincent \surnamestart Kalwa\surnameend} \&
  \bibinfo{author}{Christian \surnamestart Neurohr\surnameend}
  (\bibinfo{year}{2025}): \emph{\bibinfo{title}{TSC2CARLA: An abstract
  scenario-based verification toolchain for automated driving systems}}.
\newblock {\slshape \bibinfo{journal}{Science of Computer Programming}}
  \bibinfo{volume}{242}, p. \bibinfo{pages}{103256},
  \doi{10.1016/j.scico.2024.103256}.
\newblock
  \urlprefix\url{https://www.sciencedirect.com/science/article/pii/S0167642324001795}.

\bibitemdeclare{misc}{carla_ros_bridge_github}
\bibitem{carla_ros_bridge_github}
\bibinfo{author}{\surnamestart {carla-simulator}\surnameend}
  (\bibinfo{year}{2025}): \emph{\bibinfo{title}{{carla\_ros\_bridge: ROS/ROS2
  Bridge for CARLA Simulator}}}.
\newblock
  \bibinfo{howpublished}{\url{https://github.com/carla-simulator/ros-bridge}}.
\newblock \bibinfo{note}{Accessed: 2025-08-12}.

\bibitemdeclare{techreport}{atr117}
\bibitem{atr117}
\bibinfo{author}{Werner \surnamestart Damm\surnameend},
  \bibinfo{author}{Stephanie \surnamestart Kemper\surnameend},
  \bibinfo{author}{Eike \surnamestart M{\"o}hlmann\surnameend},
  \bibinfo{author}{Thomas \surnamestart Peikenkamp\surnameend} \&
  \bibinfo{author}{Astrid \surnamestart Rakow\surnameend}
  (\bibinfo{year}{2017}): \emph{\bibinfo{title}{{Traffic Sequence Charts - From
  Visualization to Semantics}}}.
\newblock \bibinfo{type}{{AVACS Technical Report (117)}},
  \doi{10.13140/RG.2.2.15190.42563}.

\bibitemdeclare{inproceedings}{Dosovitskiy2017}
\bibitem{Dosovitskiy2017}
\bibinfo{author}{Alexey \surnamestart Dosovitskiy\surnameend},
  \bibinfo{author}{Germ{\'a}n \surnamestart Ros\surnameend},
  \bibinfo{author}{Felipe \surnamestart Codevilla\surnameend},
  \bibinfo{author}{Antonio~M. \surnamestart L{\'o}pez\surnameend} \&
  \bibinfo{author}{Vladlen \surnamestart Koltun\surnameend}
  (\bibinfo{year}{2017}): \emph{\bibinfo{title}{CARLA: An Open Urban Driving
  Simulator}}.
\newblock In: {\slshape \bibinfo{booktitle}{Conference on Robot Learning}}.
\newblock \urlprefix\url{https://api.semanticscholar.org/CorpusID:5550767}.

\bibitemdeclare{article}{Du2019}
\bibitem{Du2019}
\bibinfo{author}{Na~\surnamestart Du\surnameend}, \bibinfo{author}{Jacob
  \surnamestart Haspiel\surnameend}, \bibinfo{author}{Qiaoning \surnamestart
  Zhang\surnameend}, \bibinfo{author}{Dawn \surnamestart Tilbury\surnameend},
  \bibinfo{author}{Anuj~K. \surnamestart Pradhan\surnameend},
  \bibinfo{author}{X.~Jessie \surnamestart Yang\surnameend} \&
  \bibinfo{author}{Lionel~P. \surnamestart Robert\surnameend}
  (\bibinfo{year}{2019}): \emph{\bibinfo{title}{Look who’s talking now:
  Implications of AV’s explanations on driver’s trust, AV preference,
  anxiety and mental workload}}.
\newblock {\slshape \bibinfo{journal}{Transportation Research Part C: Emerging
  Technologies}} \bibinfo{volume}{104}, pp. \bibinfo{pages}{428--442},
  \doi{10.1016/j.trc.2019.05.025}.

\bibitemdeclare{inproceedings}{Esterle2020}
\bibitem{Esterle2020}
\bibinfo{author}{Klemens \surnamestart Esterle\surnameend},
  \bibinfo{author}{Luis \surnamestart Gressenbuch\surnameend} \&
  \bibinfo{author}{Alois \surnamestart Knoll\surnameend}
  (\bibinfo{year}{2020}): \emph{\bibinfo{title}{Formalizing Traffic Rules for
  Machine Interpretability}}.
\newblock In: {\slshape \bibinfo{booktitle}{2020 IEEE 3rd Connected and
  Automated Vehicles Symposium (CAVS)}}, pp. \bibinfo{pages}{1--7},
  \doi{10.1109/CAVS51000.2020.9334599}.

\bibitemdeclare{article}{Finkbeiner2022}
\bibitem{Finkbeiner2022}
\bibinfo{author}{Bernd \surnamestart Finkbeiner\surnameend},
  \bibinfo{author}{Martin \surnamestart Fränzle\surnameend},
  \bibinfo{author}{Florian \surnamestart Kohn\surnameend} \&
  \bibinfo{author}{Paul \surnamestart Kröger\surnameend}
  (\bibinfo{year}{2022}): \emph{\bibinfo{title}{A Truly Robust Signal Temporal
  Logic: Monitoring Safety Properties of Interacting Cyber-Physical Systems
  under Uncertain Observation}}.
\newblock {\slshape \bibinfo{journal}{Algorithms}}
  \bibinfo{volume}{15}(\bibinfo{number}{4}), \doi{10.3390/a15040126}.

\bibitemdeclare{inproceedings}{Grundt2022}
\bibitem{Grundt2022}
\bibinfo{author}{Dominik \surnamestart Grundt\surnameend},
  \bibinfo{author}{Anna \surnamestart K\"ohne\surnameend},
  \bibinfo{author}{Ishan \surnamestart Saxena\surnameend},
  \bibinfo{author}{Ralf \surnamestart Stemmer\surnameend},
  \bibinfo{author}{Bernd \surnamestart Westphal\surnameend} \&
  \bibinfo{author}{Eike \surnamestart M\"ohlmann\surnameend}
  (\bibinfo{year}{2022}): \emph{\bibinfo{title}{Towards Runtime Monitoring of
  Complex System Requirements for Autonomous Driving Functions}}.
\newblock In \bibinfo{editor}{Matt \surnamestart Luckcuck\surnameend} \&
  \bibinfo{editor}{Marie \surnamestart Farrell\surnameend}, editors: {\slshape
  \bibinfo{booktitle}{{\rm Proceedings Fourth International Workshop on} Formal
  Methods for Autonomous Systems (FMAS) {\rm and Fourth International Workshop
  on} Automated and verifiable Software sYstem DEvelopment (ASYDE), {\rm
  Berlin, Germany, 26th and 27th of September 2022}}}, {\slshape
  \bibinfo{series}{Electronic Proceedings in Theoretical Computer Science}}
  \bibinfo{volume}{371}, \bibinfo{publisher}{Open Publishing Association}, pp.
  \bibinfo{pages}{53--61}, \doi{10.4204/EPTCS.371.4}.

\bibitemdeclare{misc}{ApplicationVideo}
\bibitem{ApplicationVideo}
\bibinfo{author}{Dominik \surnamestart Grundt\surnameend},
  \bibinfo{author}{Ishan \surnamestart Saxena\surnameend},
  \bibinfo{author}{Malte \surnamestart Petersen\surnameend},
  \bibinfo{author}{Bernd \surnamestart Westphal\surnameend} \&
  \bibinfo{author}{Eike \surnamestart Möhlmann\surnameend}
  (\bibinfo{year}{2025}): \emph{\bibinfo{title}{Application Video: Simulated
  Overtaking}}.
\newblock
  \bibinfo{howpublished}{\url{https://www.youtube.com/watch?v=5rFGkbvKOjs}}.
\newblock \bibinfo{note}{Supplementary video material illustrating the
  demonstration of our approach in a simulated overtaking}.

\bibitemdeclare{inproceedings}{Havelund2002}
\bibitem{Havelund2002}
\bibinfo{author}{Klaus \surnamestart Havelund\surnameend} \&
  \bibinfo{author}{Grigore \surnamestart Ro{\c{s}}u\surnameend}
  (\bibinfo{year}{2002}): \emph{\bibinfo{title}{Synthesizing Monitors for
  Safety Properties}}.
\newblock In \bibinfo{editor}{Joost-Pieter \surnamestart Katoen\surnameend} \&
  \bibinfo{editor}{Perdita \surnamestart Stevens\surnameend}, editors:
  {\slshape \bibinfo{booktitle}{Tools and Algorithms for the Construction and
  Analysis of Systems}}, \bibinfo{publisher}{Springer Berlin Heidelberg},
  \bibinfo{address}{Berlin, Heidelberg}, pp. \bibinfo{pages}{342--356},
  \doi{10.1007/3-540-46002-0_24}.

\bibitemdeclare{inproceedings}{Kaufman2025}
\bibitem{Kaufman2025}
\bibinfo{author}{Robert~A \surnamestart Kaufman\surnameend},
  \bibinfo{author}{Aaron \surnamestart Broukhim\surnameend},
  \bibinfo{author}{David \surnamestart Kirsh\surnameend} \&
  \bibinfo{author}{Nadir \surnamestart Weibel\surnameend}
  (\bibinfo{year}{2025}): \emph{\bibinfo{title}{What Did My Car Say? Impact of
  Autonomous Vehicle Explanation Errors and Driving Context On Comfort,
  Reliance, Satisfaction, and Driving Confidence}}.
\newblock In: {\slshape \bibinfo{booktitle}{Proceedings of the 2025 CHI
  Conference on Human Factors in Computing Systems}}, \bibinfo{series}{CHI
  '25}, \bibinfo{publisher}{Association for Computing Machinery},
  \bibinfo{address}{New York, NY, USA}, \doi{10.1145/3706598.3713088}.

\bibitemdeclare{inproceedings}{Kohl2019}
\bibitem{Kohl2019}
\bibinfo{author}{Maximilian~A. \surnamestart Köhl\surnameend},
  \bibinfo{author}{Kevin \surnamestart Baum\surnameend},
  \bibinfo{author}{Markus \surnamestart Langer\surnameend},
  \bibinfo{author}{Daniel \surnamestart Oster\surnameend},
  \bibinfo{author}{Timo \surnamestart Speith\surnameend} \&
  \bibinfo{author}{Dimitri \surnamestart Bohlender\surnameend}
  (\bibinfo{year}{2019}): \emph{\bibinfo{title}{Explainability as a
  Non-Functional Requirement}}.
\newblock In: {\slshape \bibinfo{booktitle}{2019 IEEE 27th International
  Requirements Engineering Conference (RE)}}, pp. \bibinfo{pages}{363--368},
  \doi{10.1109/RE.2019.00046}.

\bibitemdeclare{misc}{Ma2024}
\bibitem{Ma2024}
\bibinfo{author}{Shuai \surnamestart Ma\surnameend} (\bibinfo{year}{2024}):
  \emph{\bibinfo{title}{Towards Human-centered Design of Explainable Artificial
  Intelligence (XAI): A Survey of Empirical Studies}}.
\newblock \eprint{2410.21183}.

\bibitemdeclare{article}{Othman2021}
\bibitem{Othman2021}
\bibinfo{author}{Kareem \surnamestart Othman\surnameend}
  (\bibinfo{year}{2021}): \emph{\bibinfo{title}{Public acceptance and
  perception of autonomous vehicles: a comprehensive review}}.
\newblock {\slshape \bibinfo{journal}{AI Ethics}}
  \bibinfo{volume}{1}(\bibinfo{number}{3}), pp. \bibinfo{pages}{355--387},
  \doi{10.1007/s43681-021-00041-8}.

\bibitemdeclare{article}{Papagni2022}
\bibitem{Papagni2022}
\bibinfo{author}{Guglielmo \surnamestart Papagni\surnameend},
  \bibinfo{author}{Jesse \surnamestart de~Pagter\surnameend},
  \bibinfo{author}{Setareh \surnamestart Zafari\surnameend},
  \bibinfo{author}{Michael \surnamestart Filzmoser\surnameend} \&
  \bibinfo{author}{Sabine~T. \surnamestart Koeszegi\surnameend}
  (\bibinfo{year}{2022}): \emph{\bibinfo{title}{Artificial agents’
  explainability to support trust: considerations on timing and context}}.
\newblock {\slshape \bibinfo{journal}{AI Soc.}}
  \bibinfo{volume}{38}(\bibinfo{number}{2}), p. \bibinfo{pages}{947–960},
  \doi{10.1007/s00146-022-01462-7}.

\bibitemdeclare{inproceedings}{Rizaldi2017}
\bibitem{Rizaldi2017}
\bibinfo{author}{Albert \surnamestart Rizaldi\surnameend},
  \bibinfo{author}{Jonas \surnamestart Keinholz\surnameend},
  \bibinfo{author}{Monika \surnamestart Huber\surnameend},
  \bibinfo{author}{Jochen \surnamestart Feldle\surnameend},
  \bibinfo{author}{Fabian \surnamestart Immler\surnameend},
  \bibinfo{author}{Matthias \surnamestart Althoff\surnameend},
  \bibinfo{author}{Eric \surnamestart Hilgendorf\surnameend} \&
  \bibinfo{author}{Tobias \surnamestart Nipkow\surnameend}
  (\bibinfo{year}{2017}): \emph{\bibinfo{title}{Formalising and Monitoring
  Traffic Rules for Autonomous Vehicles in Isabelle/HOL}}.
\newblock In \bibinfo{editor}{Nadia \surnamestart Polikarpova\surnameend} \&
  \bibinfo{editor}{Steve \surnamestart Schneider\surnameend}, editors:
  {\slshape \bibinfo{booktitle}{Integrated Formal Methods}},
  \bibinfo{publisher}{Springer International Publishing},
  \bibinfo{address}{Cham}, pp. \bibinfo{pages}{50--66},
  \doi{10.1007/978-3-319-66845-1_4}.

\bibitemdeclare{inproceedings}{Schwammberger2025}
\bibitem{Schwammberger2025}
\bibinfo{author}{Maike \surnamestart Schwammberger\surnameend}
  (\bibinfo{year}{2025}): \emph{\bibinfo{title}{From Explanation Correctness to
  Explanation Goodness: Only Provably Correct Explanations Can Save the
  World}}.
\newblock In \bibinfo{editor}{Bernhard \surnamestart Steffen\surnameend},
  editor: {\slshape \bibinfo{booktitle}{Bridging the Gap Between AI and
  Reality}}, \bibinfo{publisher}{Springer Nature Switzerland},
  \bibinfo{address}{Cham}, pp. \bibinfo{pages}{307--317},
  \doi{10.1007/978-3-031-73741-1_19}.

\bibitemdeclare{inproceedings}{Schwammberger2022EM}
\bibitem{Schwammberger2022EM}
\bibinfo{author}{Maike \surnamestart Schwammberger\surnameend} \&
  \bibinfo{author}{Verena \surnamestart Kl\"os\surnameend}
  (\bibinfo{year}{2022}): \emph{\bibinfo{title}{From Specification Models to
  Explanation Models: An Extraction and Refinement Process for Timed
  Automata}}.
\newblock In \bibinfo{editor}{Matt \surnamestart Luckcuck\surnameend} \&
  \bibinfo{editor}{Marie \surnamestart Farrell\surnameend}, editors: {\slshape
  \bibinfo{booktitle}{{\rm Proceedings Fourth International Workshop on} Formal
  Methods for Autonomous Systems (FMAS) {\rm and Fourth International Workshop
  on} Automated and verifiable Software sYstem DEvelopment (ASYDE), {\rm
  Berlin, Germany, 26th and 27th of September 2022}}}, {\slshape
  \bibinfo{series}{Electronic Proceedings in Theoretical Computer Science}}
  \bibinfo{volume}{371}, \bibinfo{publisher}{Open Publishing Association}, pp.
  \bibinfo{pages}{20--37}, \doi{10.4204/EPTCS.371.2}.

\bibitemdeclare{inproceedings}{Schwammberger2024xAI}
\bibitem{Schwammberger2024xAI}
\bibinfo{author}{Maike \surnamestart Schwammberger\surnameend},
  \bibinfo{author}{Raffaela \surnamestart Mirandola\surnameend} \&
  \bibinfo{author}{Nils \surnamestart Wenninghoff\surnameend}
  (\bibinfo{year}{2024}): \emph{\bibinfo{title}{Explainability Engineering
  Challenges: Connecting Explainability Levels to Run-Time Explainability}}.
\newblock In \bibinfo{editor}{Luca \surnamestart Longo\surnameend},
  \bibinfo{editor}{Sebastian \surnamestart Lapuschkin\surnameend} \&
  \bibinfo{editor}{Christin \surnamestart Seifert\surnameend}, editors:
  {\slshape \bibinfo{booktitle}{Explainable Artificial Intelligence}},
  \bibinfo{publisher}{Springer Nature Switzerland}, \bibinfo{address}{Cham},
  pp. \bibinfo{pages}{205--218}, \doi{10.1007/978-3-031-63803-9_11}.

\bibitemdeclare{article}{Stemmer2025}
\bibitem{Stemmer2025}
\bibinfo{author}{Ralf \surnamestart Stemmer\surnameend}, \bibinfo{author}{Ishan
  \surnamestart Saxena\surnameend}, \bibinfo{author}{Lukas \surnamestart
  Panneke\surnameend}, \bibinfo{author}{Dominik \surnamestart
  Grundt\surnameend}, \bibinfo{author}{Anna \surnamestart Austel\surnameend},
  \bibinfo{author}{Eike \surnamestart Möhlmann\surnameend} \&
  \bibinfo{author}{Bernd \surnamestart Westphal\surnameend}
  (\bibinfo{year}{2025}): \emph{\bibinfo{title}{Runtime monitoring of complex
  scenario-based requirements for autonomous driving functions}}.
\newblock {\slshape \bibinfo{journal}{Science of Computer Programming}}
  \bibinfo{volume}{244}, p. \bibinfo{pages}{103301},
  \doi{10.1016/j.scico.2025.103301}.

\bibitemdeclare{inproceedings}{Tuncali2018}
\bibitem{Tuncali2018}
\bibinfo{author}{Cumhur~Erkan \surnamestart Tuncali\surnameend},
  \bibinfo{author}{Georgios \surnamestart Fainekos\surnameend},
  \bibinfo{author}{Hisahiro \surnamestart Ito\surnameend} \&
  \bibinfo{author}{James \surnamestart Kapinski\surnameend}
  (\bibinfo{year}{2018}): \emph{\bibinfo{title}{Simulation-based Adversarial
  Test Generation for Autonomous Vehicles with Machine Learning Components}}.
\newblock In: {\slshape \bibinfo{booktitle}{2018 IEEE Intelligent Vehicles
  Symposium (IV)}}, pp. \bibinfo{pages}{1555--1562},
  \doi{10.1109/IVS.2018.8500421}.

\end{thebibliography}
\newpage
\section*{Annex}
\label{sec:annex}

\begin{table}[H]
	\centering
	\caption{Specified (un)expectable driving manoeuvres ($\FEi{1}, \FEi{2},\FVi{1}$) and their corresponding predicate logical formulas for \textit{Phase 1 - Approaching} of the overtaking scenario. The manoeuvres are represented as abstract traffic scenarios using TSC~\cite{atr117}, according to the object model $OM$ \autoref{fig:exOM} and symbol dictionary \autoref{fig:exSD} (except that we added $indicators_{left}$ attribute for class \textit{Car}, and a visual feature (yellow blobs) in the symbol dictionary for the indicators). The corresponding context $C_1$ and formula are introduced in \autoref{fig:eval-phase1-sv}. The semicolon separates the invariant nodes, indicating the temporal partitioning of the time interval $[b,e]$ over which the sequence chart is evaluated.}
	\renewcommand{\arraystretch}{1.4} 
	\setlength{\tabcolsep}{5pt}       
	\begin{tabular}{|>{\centering\arraybackslash}m{0.46\textwidth} 
			|>{\centering\arraybackslash}m{0.5\textwidth}|}
		\hline
		\textbf{Specified Manoeuvrers} & \textbf{Predicate Logical Formula} \\
		\hline
		\includegraphics[width=\linewidth]{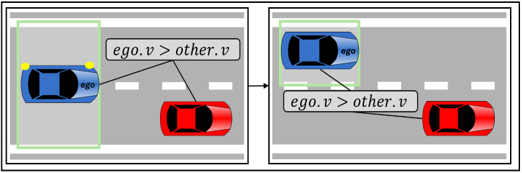} &
		\(\displaystyle
		\begin{array}{l}
			F_{E_1} = ego.pos.y > lane.yR \land ego.pos.y < lane.yL \\
			\quad \land\ other.pos.y > lane.yR \land other.pos.y < lane.yM \\
			\quad \land\ ego.pos.x < other.pos.x \\
			\quad \land\ ego.v > other.v \\
			\quad \land\ ego.indicators_{left} = 1; \\
			\quad ego.pos.y > lane.yM \land ego.pos.y < lane.yL \\
			\quad \land\ other.pos.y > lane.yR \land other.pos.y < lane.yM \\
			\quad \land\ ego.pos.x < other.pos.x \\
			\quad \land\ ego.v > other.v
		\end{array}
		\)
		 \\
		\hline
		\includegraphics[width=\linewidth]{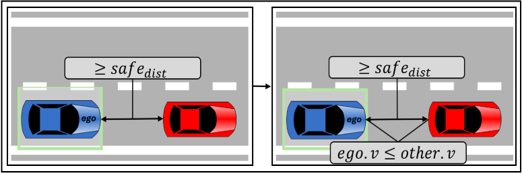} &
		\(\displaystyle
		\begin{array}{l}
			F_{E_2} = ego.pos.y > lane.yR \land ego.pos.y < lane.yM \\
			\quad \land\ other.pos.y > lane.yR \land other.pos.y < lane.yM \\
			\quad \land\ ego.pos.x + safe_{dist} < other.pos.x; \\
			\quad ego.pos.y > lane.yR \land ego.pos.y < lane.yM \\
			\quad \land\ other.pos.y > lane.yR \land other.pos.y < lane.yM \\
			\quad \land\ ego.pos.x + safe_{dist} < other.pos.x \\
			\quad \land\ ego.v \leq other.v
		\end{array}
		\)
		 \\
		\hline
		\includegraphics[width=\linewidth]{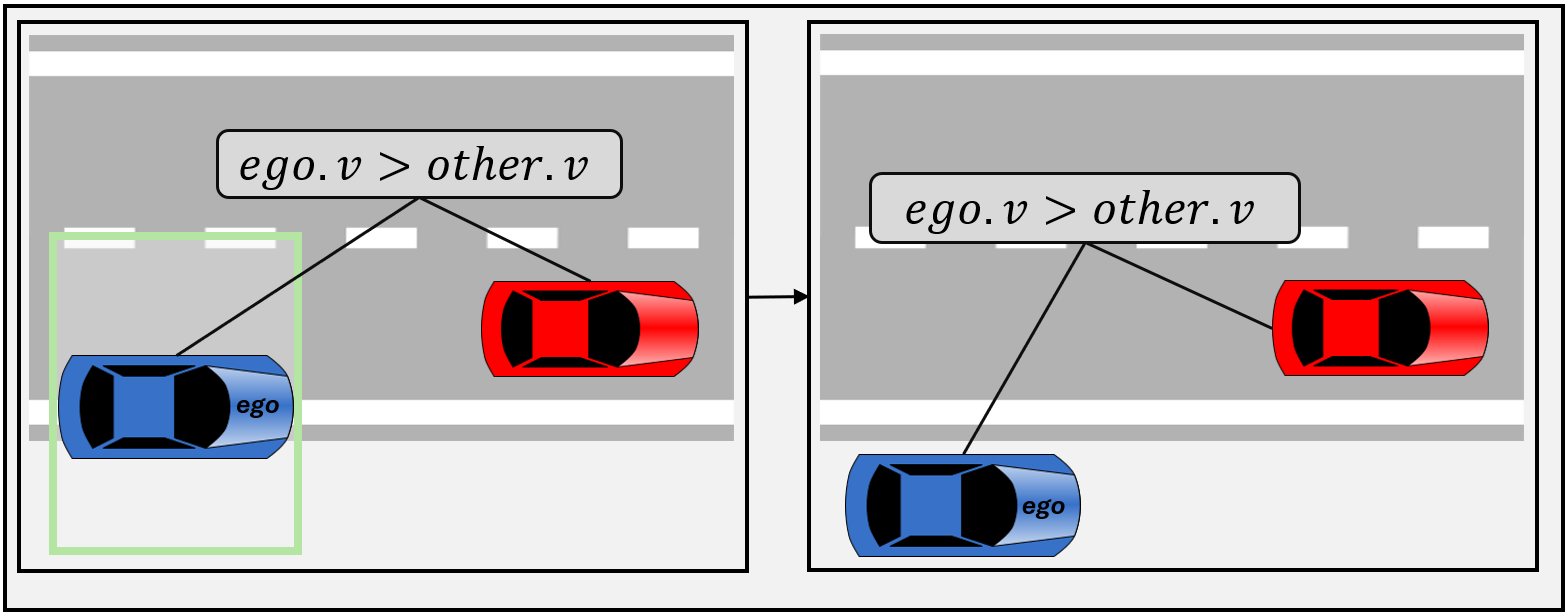} &
		\(\displaystyle
		\begin{array}{l}
			F_{V_1} = ego.pos.y > lane.yM \\
			\quad \land\ other.pos.y > lane.yR \land other.pos.y < lane.yM \\
			\quad \land\ ego.pos.x < other.pos.x \\
			\quad \land\ ego.v > other.v; \\
			\quad ego.pos.y < lane.yR \\
			\quad \land\ other.pos.y > lane.yR \land other.pos.y < lane.yM \\
			\quad \land\ ego.pos.x < other.pos.x \\
			\quad \land\ ego.v > other.v
		\end{array}
		\)
		 \\
		\hline
	\end{tabular} 
	\label{tab:phase1-manoeuvres}
\end{table}

\begin{table}[t]
	\centering
	\caption{Specified (un)expectable driving manoeuvres ($\FEi{1}, \FVi{1}$) and their corresponding predicate logical formulas for \textit{Phase 2 - Closing Gap} of the overtaking scenario. The manoeuvres are represented as abstract traffic scenarios using TSC~\cite{atr117}, according to the object model $OM$ \autoref{fig:exOM} and symbol dictionary \autoref{fig:exSD} (except that we added $indicators_{left}$ attribute for class \textit{Car}, and a visual feature (yellow blobs) in the symbol dictionary for the indicators). The corresponding context $C_2$ and formula are introduced in \autoref{eq:exsv}. The semicolon separates the invariant nodes, indicating the temporal partitioning of the time interval $[b,e]$ over which the sequence chart is evaluated.}
	\renewcommand{\arraystretch}{1.4} 
	\setlength{\tabcolsep}{5pt}       
	\begin{tabular}{|>{\centering\arraybackslash}m{0.46\textwidth} 
			|>{\centering\arraybackslash}m{0.5\textwidth}|}
		\hline
		\textbf{Specified Manoeuvrers} & \textbf{Predicate Logical Formula} \\
		\hline
		\includegraphics[width=\linewidth]{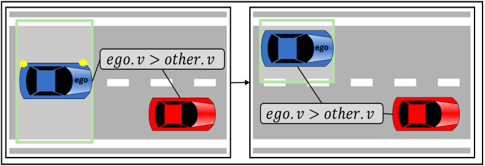} &
		\(\displaystyle
		\begin{array}{l}
			F_{E_1} = ego.pos.y > lane.yR \land ego.pos.y < lane.yL \\
			\quad \land\ other.pos.y > lane.yR \land other.pos.y < lane.yM \\
			\quad \land\ ego.pos.x < other.pos.x \\
			\quad \land\ ego.v > other.v \\
			\quad \land\ ego.indicators_{left} = 1; \\
			\quad ego.pos.y > lane.yM \land ego.pos.y < lane.yL \\
			\quad \land\ other.pos.y > lane.yR \land other.pos.y < lane.yM \\
			\quad \land\ ego.pos.x < other.pos.x \\
			\quad \land\ ego.v > other.v
		\end{array}
		\)
		\\
		\hline
		\includegraphics[width=\linewidth]{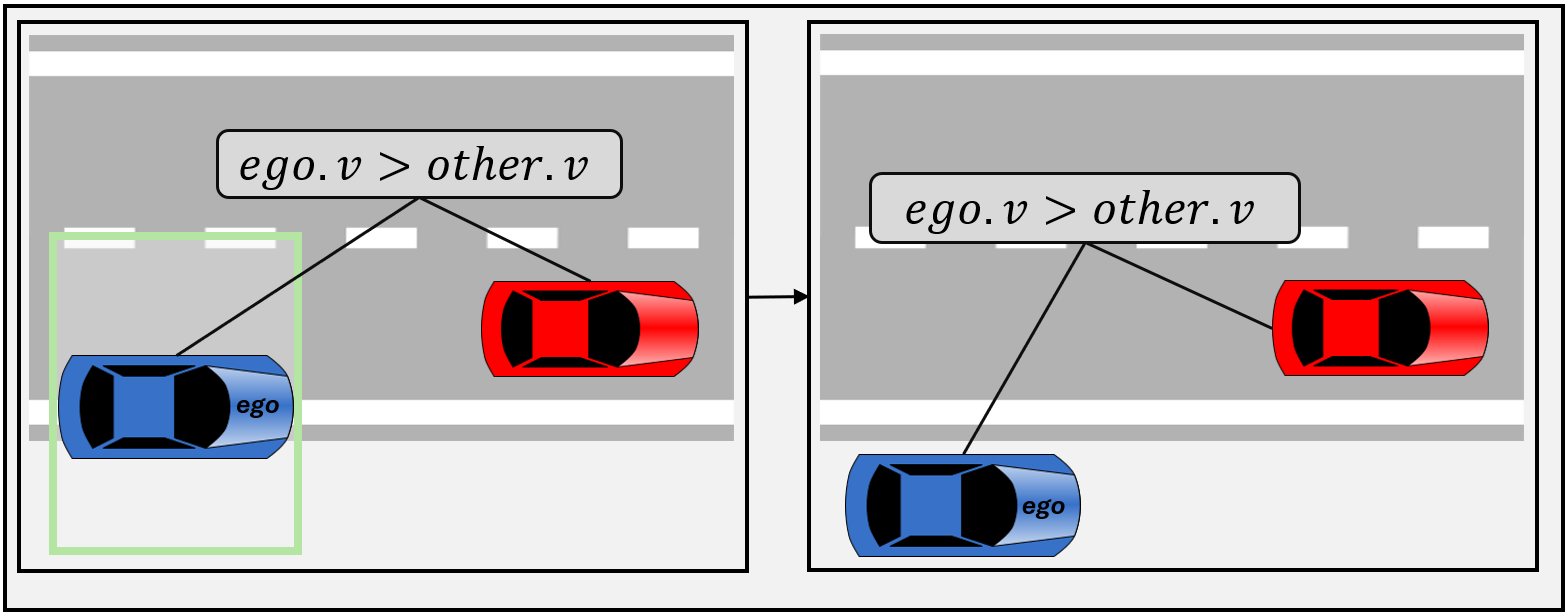} &
		\(\displaystyle
		\begin{array}{l}
			F_{V_1} = ego.pos.y < lane.yM \\
			\quad \land\ other.pos.y > lane.yR \land other.pos.y < lane.yM \\
			\quad \land\ ego.pos.x < other.pos.x \\
			\quad \land\ ego.v > other.v; \\
			\quad ego.pos.y < lane.yR \\
			\quad \land\ other.pos.y > lane.yR \land other.pos.y < lane.yM \\
			\quad \land\ ego.pos.x < other.pos.x \\
			\quad \land\ ego.v > other.v
		\end{array}
		\)
		\\
		\hline
	\end{tabular}
	\label{tab:phase2-manoeuvres}
\end{table}

\begin{table}[t]
	\centering
	\caption{Specified context $C_3$ and (un)expectable driving manoeuvres ($\FEi{1}, \FVi{1}$), and corresponding predicate logical formulas for \textit{Phase 3 - Lane Change} of the overtaking scenario. The specifications are based on the object model $OM$ \autoref{fig:exOM} and symbol dictionary \autoref{fig:exSD} (except that we added $indicators_{left}$ attribute for class \textit{Car}, and a visual feature (yellow blobs) in the symbol dictionary for the indicators). The semicolon separates the invariant nodes, indicating the temporal partitioning of the time interval $[b,e]$ over which the sequence chart is evaluated.}
	\renewcommand{\arraystretch}{1.4} 
	\setlength{\tabcolsep}{5pt}       
	\begin{tabular}{|>{\centering\arraybackslash}m{0.46\textwidth} 
			|>{\centering\arraybackslash}m{0.5\textwidth}|}
		\hline
		\textbf{Specified Context \& Manoeuvrers} & \textbf{Predicate Logical Formula} \\
		\hline
		\includegraphics[scale=0.2]{figures/eval-phase3-sv.png} &
		\(\displaystyle
		\begin{array}{l}
			C_3 = ego.pos.y > lane.yR \land ego.pos.y < lane.yL \\
			\quad \land\ other.pos.y > lane.yR \land other.pos.y < lane.yM \\
			\quad \land\ ego.pos.x  < other.pos.x \\
			\quad \land\ ego.indicators_{left} = 1
		\end{array}
		\)
		
		\\
		\hline
		\includegraphics[width=\linewidth]{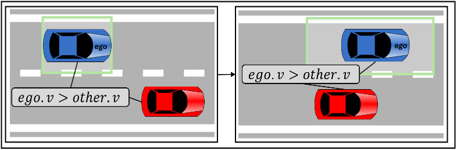} &
		\(\displaystyle
		\begin{array}{l}
			F_{E_1} = ego.pos.y > lane.yM \land ego.pos.y < lane.yL \\
			\quad \land\ other.pos.y > lane.yR \land other.pos.y < lane.yM \\
			\quad \land\ ego.pos.x < other.pos.x \\
			\quad \land\ ego.v > other.v; \\
			\quad ego.pos.y > lane.yM \land ego.pos.y < lane.yL \\
			\quad \land\ other.pos.y > lane.yR \land other.pos.y < lane.yM \\
			\quad \land\ ego.pos.x \geq other.pos.x \\
			\quad \land\ ego.v > other.v
		\end{array}
		\)
		\\
		\hline
		\includegraphics[width=\linewidth]{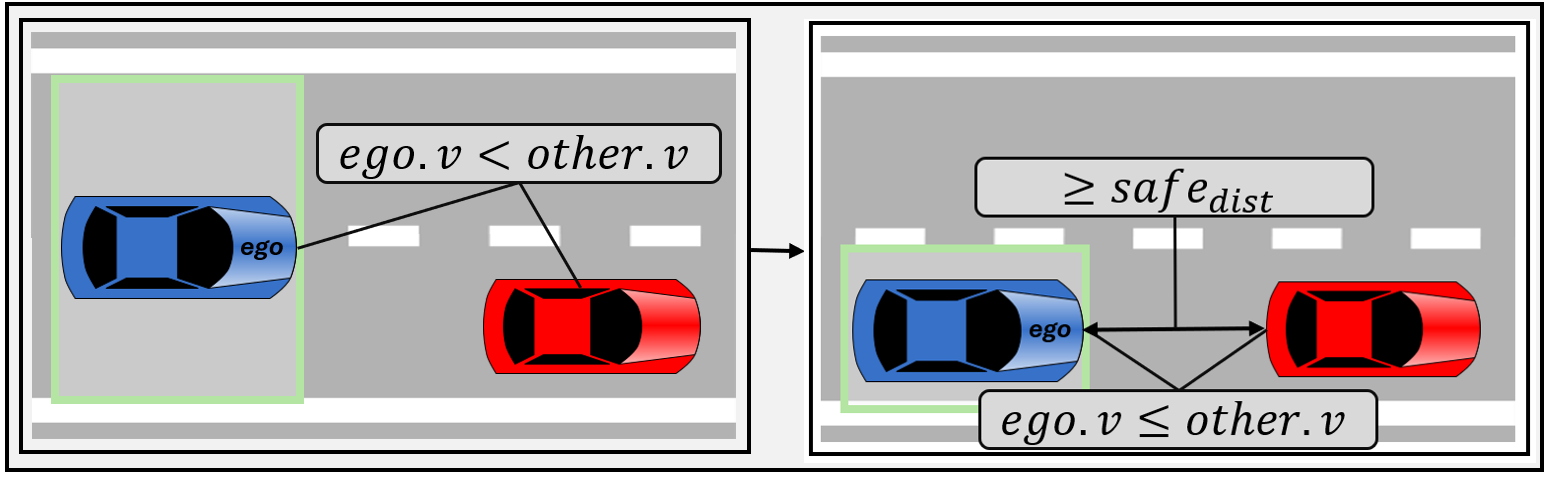} &
		\(\displaystyle
		\begin{array}{l}
			F_{V_1} = ego.pos.y > lane.yR \land ego.pos.y < lane.yL \\
			\quad \land\ other.pos.y > lane.yR \land other.pos.y < lane.yM \\
			\quad \land\ ego.pos.x < other.pos.x \\
			\quad \land\ ego.v < other.v; \\
			\quad ego.pos.y > lane.yR \land ego.pos.y < lane.yM \\
			\quad \land\ other.pos.y > lane.yR \land other.pos.y < lane.yM \\
			\quad \land\ ego.pos.x + safe_{dist} \leq other.pos.x \\
			\quad \land\ ego.v \leq other.v
		\end{array}
		\)
		\\
		\hline
	\end{tabular}
	\label{tab:phase3-manoeuvres}
\end{table}

\begin{table}[t]
	\centering
	\caption{Specified context $C_4$ and (un)expectable driving manoeuvres ($\FEi{1}, \FVi{1}$), and corresponding predicate logical formulas for \textit{Phase 4 - Overtaking} of the overtaking scenario. The specifications are based on the object model $OM$ \autoref{fig:exOM} and symbol dictionary \autoref{fig:exSD} (except that we added $indicators_{left}$ attribute for class \textit{Car}, and a visual feature (yellow blobs) in the symbol dictionary for the indicators). The semicolon separates the invariant nodes, indicating the temporal partitioning of the time interval $[b,e]$ over which the sequence chart is evaluated.}
	\renewcommand{\arraystretch}{1.4} 
	\setlength{\tabcolsep}{5pt}       
	\begin{tabular}{|>{\centering\arraybackslash}m{0.46\textwidth} 
			|>{\centering\arraybackslash}m{0.5\textwidth}|}
		\hline
		\textbf{Specified Context \& Manoeuvrers} & \textbf{Predicate Logical Formula} \\
			\hline
		\includegraphics[scale=0.2]{figures/eval-phase4-sv.png} &
		\(\displaystyle
		\begin{array}{l}
			C_4 = ego.pos.y > lane.yM \land ego.pos.y < lane.yL \\
			\quad \land\ other.pos.y > lane.yR \land other.pos.y < lane.yM \\
			\quad \land\ ego.pos.x \leq other.pos.x \\
			\quad \land\ ego.v > other.v
		\end{array}
		\)
		\\
		\hline
		\includegraphics[width=\linewidth]{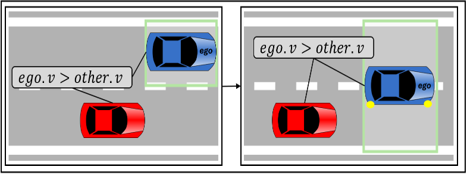} &
		\(\displaystyle
		\begin{array}{l}
			F_{E_1} = ego.pos.y > lane.yM \land ego.pos.y < lane.yL \\
			\quad \land\ other.pos.y > lane.yR \land other.pos.y < lane.yM \\
			\quad \land\ ego.pos.x > other.pos.x \\
			\quad \land\ ego.v > other.v;\\
			\quad ego.pos.y > lane.yR \land ego.pos.y < lane.yL \\
			\quad \land\ other.pos.y > lane.yR \land other.pos.y < lane.yM \\
			\quad \land\ ego.pos.x > other.pos.x \\
			\quad \land\ ego.v > other.v\\
			\quad \land\ ego.indicators_{right} = 1
		\end{array}
		\)
		\\
		\hline
		\includegraphics[width=\linewidth]{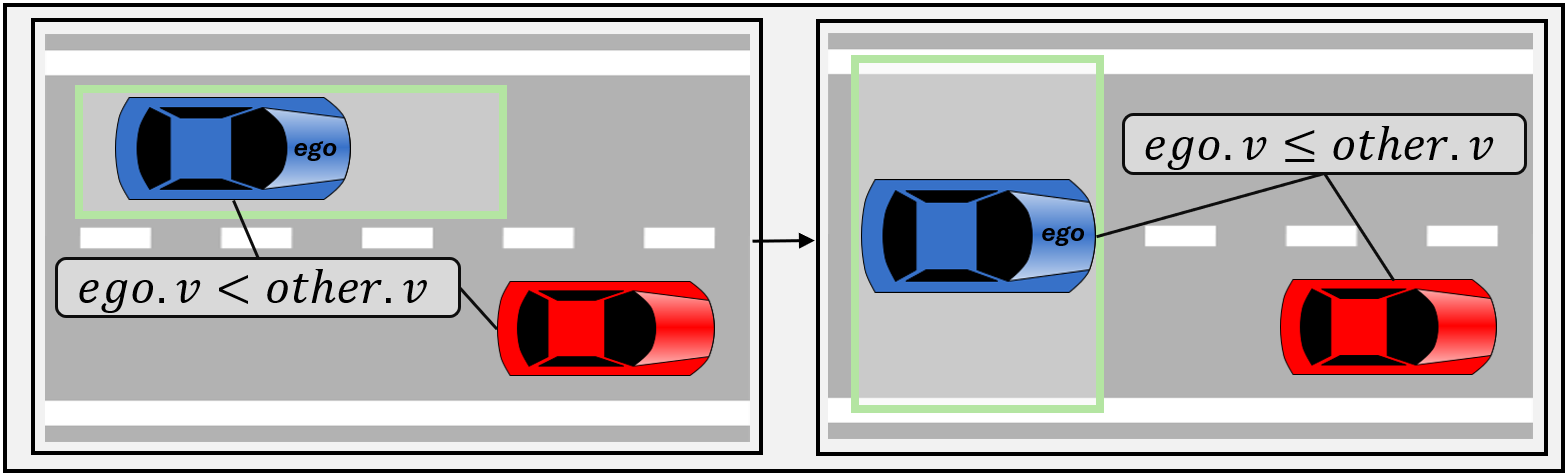} &
		\(\displaystyle
		\begin{array}{l}
			F_{V_1} = ego.pos.y > lane.yM \land ego.pos.y < lane.yL \\
			\quad \land\ other.pos.y > lane.yR \land other.pos.y < lane.yM \\
			\quad \land\ ego.pos.x < other.pos.x \\
			\quad \land\ ego.v < other.v; \\
			\quad ego.pos.y > lane.yR \land ego.pos.y < lane.yL \\
			\quad \land\ other.pos.y > lane.yR \land other.pos.y < lane.yM \\
			\quad \land\ ego.pos.x < other.pos.x \\
			\quad \land\ ego.v \leq other.v \\
		\end{array}
		\)
		\\
		\hline
	\end{tabular}
	\label{tab:phase4-manoeuvres}
\end{table}

\end{document}